\documentclass[prd,aps,preprint,showpacs,nofootinbib,superscriptaddress]{revtex4}
\usepackage[latin1]{inputenc}
\usepackage{epsfig}
\usepackage{amsmath}
\usepackage{latexsym}
\usepackage{amssymb}
\usepackage{dsfont}
\usepackage{color}
\usepackage{multirow}
\usepackage{enumitem}

\newcommand{\ud}{\mathrm{d}}

\newcommand{\uM}{\mathcal{M}}

\newcommand{\uTr}{\mathrm{Tr}}

\newcommand{\uslash}{/\!\!\!}

\newcommand{\usigma}{\boldsymbol{\sigma}}

\newcommand{\upi}{\boldsymbol{\pi}}

\newcommand{\uS}{\boldsymbol{S}}
\newcommand{\us}{\boldsymbol{s}}
\newcommand{\uk}{\boldsymbol{k}}

\newcommand{\uzero}{\boldsymbol{0}}

\newlength\savedwidth

\begin{document}

\newcommand*{\Mainz}{Institut für Kernphysik, Johannes Gutenberg-Universität,\\  D-55099 Mainz, Germany}\affiliation{\Mainz}
\newcommand*{\Pavia}
{
Dipartimento di Fisica Nucleare e Teorica, Universit\`a degli Studi di Pavia, \\and INFN, Sezione di Pavia, I-27100 Pavia, Italy}\affiliation{\Pavia}

\title{On the Origin of Model Relations among  Transverse-Momentum Dependent Parton Distributions}

\author{C. Lorc\'e\footnote{E-mail: lorce@kph.uni-mainz.de}}\affiliation{\Mainz}
\author{
B. Pasquini\footnote{E-mail: pasquini@pv.infn.it}}\affiliation{\Pavia}

\begin{abstract}
{
Transverse-momentum dependent parton distributions (TMDs) 
are studied in the framework of quark models. In particular, 
 quark-model relations among TMDs are reviewed, 
elucidating their physical origin in terms of the 
quark-spin structure in the nucleon.
The formal aspects of the derivation of these relations are 
complemented with explicit examples, emphasizing how and to which extent 
the conditions which lead to relations among TMDs are  implemented in 
different classes of quark models.
}

\end{abstract}


\maketitle

\section{Introduction}
\label{section-1}

Transverse-momentum dependent parton distributions (TMDs) have received a great attention in the last years as 
they represent key objects to map out the three-dimensional partonic structure of hadrons 
in momentum space. 
The dependence on the transverse momentum of the quark 
allows for a full account of the orbital motion of the quarks 
and introduces non-trivial correlations between the orbital angular momentum
and the spin of the quark inside nucleons with different polarization states.
TMDs typically give rise to spin and azimuthal asymmetries in, for instance, semi-inclusive deep inelastic scattering and Drell-Yan processes, and significant efforts have already been devoted to measure these observables (see, e.g., Ref.~\cite{Barone:2010zz} for a recent review).
However, the extraction of TMDs from experimental data is a quite 
difficult task and needs educated Ans\"atze for fits of TMD parametrizations. 
To this aim, model calculations of TMDs play a crucial role and 
are essential towards an understanding of the non-perturbative aspects
of TMDs.
At leading twist there are eight TMDs, three of them surviving when integrated over the transverse momentum and giving rise to the familiar parton density, helicity and transversity distributions.
Studies of the TMDs have been  mainly focused on the quark contribution, 
and predictions have been obtained within a variety of models~\cite{Jakob:1997wg,Goldstein:2002vv, Pobylitsa:2002fr,Gamberg:2003ey,Lu:2004au,Cherednikov:2006zn,Lu:2006kt,Gamberg:2007wm,Pasquini:2009bv,Efremov:2009ze,She:2009jq,Zhu:2011zza,Lu:2010dt,Avakian:2009jt,Courtoy:2008dn,Courtoy:2009pc,Courtoy:2008vi,Pasquini:2010af,Brodsky:2010vs,Lorce:2007fa,Pasquini:2008ax,Boffi:2009sh,Pasquini:2010pa,Lorce:2011dv,Bacchetta:2008af,Avakian:2010br,Avakian:2008dz,Efremov:2010mt,Meissner:2007rx,Wakamatsu:2009fn,Schweitzer:2001sr}.
Despite the specific assumptions for modeling the quark dynamics, 
most of these models predicted relations among the 
leading-twist TMDs. Since in QCD the eight TMDs are all independent, it is clear that such relations should be traced back to some common simplifying 
assumptions in the models. First of all, it was noticed that they break down in models with gauge-field degrees of freedom. Furthermore, 
most  quark models are valid at some very low scale and these relations
 are expected to break under QCD evolution  to higher scales.
Despite these limitations, such relations are intriguing
 because they can provide guidelines for building parametrizations of TMDs to be tested with experimental data and also  can give useful insights  for the understanding of the origin 
of the different spin-orbit correlations of quarks in the nucleon.

The aim of this paper is to review these model relations and in particular to 
explain their physical origin.
In Sec. II we give the formalism for the definition of the leading-twist TMDs, and 
introduce a convenient representation of the quark-quark correlator in terms of the net  polarization states of the quark and hadrons.
The model relations among TMDs are introduced and explained in details in Sec. III. In particular, there are two linear relations and a quadratic relation which are flavor independent and involve polarized TMDs, while a further linear relation is flavor dependent and involves both polarized and unpolarized TMDs.

The relations among polarized TMDs connect the distributions 
of quarks inside the nucleon for different configurations of the polarization states of the hadron and the partons. As a consequence, it is natural to expect that they can originate
from rotational invariance of the polarization states of the system.
Rotations are more easily discussed in the basis of canonical spin. 
Therefore, instead of working in the standard basis of light-cone helicity, 
we  introduce in Secs. IV-A and IV-B the tensor correlator defining 
the TMDs in the canonical-spin basis. Such a representation  is used in 
Sec. IV-C to discuss 
the consequences of rotational symmetries of the system.
In such a way we will be able to identify the key ingredients for the existence
of relations among polarized TMDs in quark models.
\\ In order to complete the discussion, including the flavor-dependent
relation among polarized and unpolarized TMDs, we need to introduce
specific assumptions 
about the spin-isospin structure
of the nucleon state.
Therefore, in Sec. V we discuss the consequences of rotational invariance 
using the explicit representation of the TMDs in terms of three-quark (3Q) wave functions.
In particular, in Sec. V-A we derive the overlap representation of the TMDs in terms of light-cone wave functions, while the corresponding representation in terms of wave function in the canonical-spin basis is given in Sec. V-B.
In Sec. V-C we discuss the constraints of rotational symmetry on the nucleon wave function and, as a result, we give an alternative derivation of the flavor-independent relations among TMDs.
Finally, in Sec.  V-D we discuss the constraint due to $SU(6)$ symmetry of
the spin-flavor dependent part of the nucleon wave function. This additional 
ingredient allows us to explain the origin of the flavor-dependent relation.

The formal derivation of the relations among TMDs is made explicit within
different quark models in the final Sec. VI.
There we review different quark models which have been used in literature 
for the calculation of TMDs. In particular, we discuss the key ingredients
of the models, showing how and to which extent the conditions which lead to relations among TMDs are realized.
A summary of our findings 
 is given in the final section.
Technical details and further explanations about the representation
in terms of nucleon wave functions are collected in three appendices.

\section{Transverse-Momentum Dependent Parton Distributions}
\label{section-2}

\subsection{Definitions}

In this section, we review the formalism for the definition of TMDs, following the conventions of Refs.~\cite{Mulders:1995dh,Boer:1997nt,Goeke:2005hb}.
Introducing two lightlike four-vectors $n_\pm$ satisfying $n_+\cdot n_-=1$, we write the light-cone components of a generic four-vector $a$ as $\left[a^+,a^-,\boldsymbol a_\perp\right]$ with $a^\pm=a\cdot n_\mp$.
\newline
\noindent
The density of quarks can be defined from the following quark-quark correlator 
\begin{equation}
\Phi_{ab}(x,\uk_\perp,S)=\int\frac{\ud\xi^-\,\ud^2\xi_\perp}{(2\pi)^3}\,e^{i\left(k^+\xi^--\uk_\perp\cdot\boldsymbol\xi_\perp\right)}\langle P,S|\overline\psi_b(0)\mathcal U^{n_-}_{(0,+\infty)}\mathcal U^{n_-}_{(+\infty,\xi)}\psi_a(\xi)|P,S\rangle\big|_{\xi^+=0},
\label{correlator}
\end{equation}
where $k^+=xP^+$, $\psi$ is the quark field operator with $a,b$ 
 indices in the Dirac space, and $\mathcal U$ is the Wilson line which ensures color gauge invariance \cite{Bomhof:2004aw}. The target state is characterized by its four-momentum $P$ and covariant spin four-vector $S$ satisfying $P^2=M^2$, $S^2=-1$, and $P\cdot S=0$. We choose a frame where the hadron momentum has no transverse components $P=\left[P^+,\tfrac{M^2}{2P^+},\uzero_\perp\right]$, and so $S=\left[S_z\,\tfrac{P^+}{M},-S_z\,\tfrac{M}{2P^+},\uS_\perp\right]$ with $\uS^2=1$. From now on, we replace the dependence on the covariant spin four-vector $S$ by the dependence on the unit three-vector $\uS=\left(\uS_\perp,S_z\right)$.

TMDs enter the general Lorentz-covariant decomposition of the correlator $\Phi_{ab}(x,\uk_\perp,\uS)$ which, at twist-two level and for a spin-$1/2$ target, reads
\begin{multline}
\Phi(x,\uk_\perp,\uS)=\frac{1}{2}\Big\{f_1\,\uslash n_+-\tfrac{\epsilon_T^{ij}\,k_\perp^iS_\perp^j}{M}\,f_{1T}^\perp\,\uslash n_++S_z\,g_{1L}\,\gamma_5\uslash n_++\tfrac{\uk_\perp\cdot\uS_\perp}{M}\,g_{1T}\,\gamma_5\uslash n_+\\
+h_{1T}\,\tfrac{[\uslash S_\perp,\uslash n_+]}{2}\,\gamma_5+S_z\,h_{1L}^\perp\,\tfrac{[\uslash k_\perp,\uslash n_+]}{2M}\,\gamma_5+\tfrac{\uk_\perp\cdot\uS_\perp}{M}\,h_{1T}^\perp\,\tfrac{[\uslash k_\perp,\uslash n_+]}{2M}\,\gamma_5+ih_1^\perp\,\tfrac{[\uslash k_\perp,\uslash n_+]}{2M}\Big\},
\end{multline}
where $\epsilon_T^{12}=-\epsilon_T^{21}=1$, and the transverse four-vectors are defined as $a_\perp=\left[0,0,\boldsymbol a_\perp\right]$. The nomenclature of the distribution functions follows closely that of  Ref.~\cite{Mulders:1995dh}, sometimes referred to as ``Amsterdam notation'': $f$ refers to unpolarized target; $g$ and $h$ to longitudinally and transversely polarized target, respectively; a subscript $1$ is given to the twist-two functions; subscripts $L$ or $T$ refer to the connection with the hadron spin being longitudinal or transverse; and a symbol $\perp$ signals the explicit presence of transverse momenta with an uncontracted index. Among these eight distributions, the so-called Boer-Mulders function $h_1^\perp$~\cite{Boer:1997nt} and Sivers function $f_{1T}^\perp$~\cite{Sivers:1989cc} are T-odd, \emph{i.e.} they change sign under ``naive time-reversal'', which is defined as usual time-reversal but without interchange of initial and final states. All the TMDs depend on $x$ and $\uk^2_\perp$. These functions can be individually isolated by performing traces of the correlator with suitable Dirac matrices. Using the abbreviation $\Phi^{[\Gamma]}\equiv\uTr[\Phi\Gamma]/2$, we have
\begin{align}
\Phi^{[\gamma^+]}(x,\uk_\perp,\uS)&=f_1-\tfrac{\epsilon_T^{ij}\,k_\perp^iS_\perp^j}{M}\,f_{1T}^\perp,
\label{vector}\\
\Phi^{[\gamma^+\gamma_5]}(x,\uk_\perp,\uS)&=S_z\,g_{1L}+\tfrac{\uk_\perp\cdot\uS_\perp}{M}\,g_{1T},\\
\Phi^{[i\sigma^{j+}\gamma_5]}(x,\uk_\perp,\uS)&=S_\perp^j\,h_1+S_z\,\tfrac{k_\perp^j}{M}\,h^\perp_{1L}+S^i_\perp\,\tfrac{2k^i_\perp k^j_\perp-\uk^2_\perp\delta^{ij}}{2M^2}\,h^\perp_{1T}+\tfrac{\epsilon_T^{ji}\,k_\perp^i}{M}\,h_1^\perp,
\label{tensor}
\end{align}
where $j=1,2$ is a transverse index, and
\begin{equation}
h_1=h_{1T}+\tfrac{\uk^2_\perp}{2M^2}\,h^\perp_{1T}.
\end{equation}

The correlation function $\Phi^{[\gamma^+]}(x,\uk_\perp,\uS)$ is just the unpolarized quark distribution, which integrated over $\uk_\perp$ gives the familiar light-cone momentum distribution $f_1(x)$. All the other TMDs characterize the strength of different spin-spin and spin-orbit correlations. The precise form of this correlation is given by the prefactors of the TMDs in Eqs.~\eqref{vector}-\eqref{tensor}. In particular, the TMDs $g_{1L}$ and $h_1$ describe the strength of a correlation between a longitudinal/transverse target polarization and a longitudinal/transverse parton polarization. After integration over $\uk_\perp$, they reduce to the helicity and transversity distributions, respectively. By definition, the spin-orbit correlations described by $f_{1T}^\perp$, $g_{1T}$, $h_1^\perp$, $h_{1L}^\perp$ and $h_{1T}^\perp$ involve the transverse parton momentum and the polarization of both the parton and the target, and vanish upon integration over $\uk_\perp$.

In the following we will focus the discussion on the quark contribution 
to TMDs, ignoring the contribution from gauge fields
and therefore reducing the gauge links in Eq.~(\ref{correlator}) 
to the identity.

\subsection{Helicity and Four-Component Bases}
\label{sect:II-B}
The physical meaning of the correlations encoded in TMDs becomes especially transparent when using for the quark fields the expansion in terms of light-cone Fock operators. We consider in this study only the positive-frequency part of the quark field. The negative-frequency, corresponding to antiquark degrees of freedom, can be treated in a similar way. Moreover, we decompose the correlator into the different quark flavor contributions
\begin{equation}
\Phi=\sum_q\Phi_q.
\end{equation} 
Following the lines of Refs.~\cite{Diehl:2000xz,Brodsky:2000xy,Pasquini:2008ax}, we obtain at twist-two level
\begin{equation}\label{corrTMD}
\Phi^{[\Gamma]}_q(x,\uk_\perp,\uS)=\frac{1}{\mathcal N}\,\langle P,\uS|\sum_{\lambda'\lambda}q^\dag_{\lambda'}(\tilde k)\,q_\lambda(\tilde k)\,M^{[\Gamma]\lambda'\lambda}|P,\uS\rangle,
\end{equation}
where $\mathcal N=\left[2x(2\pi)^3\right]^2\delta^{(3)}(\uzero)$ and $M^{[\Gamma]\lambda'\lambda}=\overline u_{LC}(k,\lambda')\Gamma u_{LC}(k,\lambda)/2k^+$ with $u_{LC}(k,\lambda)$ the free Dirac light-cone spinor (see App.~\ref{appendix:a}). The operators $q^\dag_\lambda(\tilde k)$ and $q_\lambda(\tilde k)$ respectively create and annihilate a quark with flavor $q$, light-cone helicity $\lambda$, and light-cone momentum $\tilde k=(xP^+,\uk_\perp)$. 

We find very convenient to associate a four-component vector\footnote{Note this is \emph{not} a Lorentz four-vector but Einstein's summation convention still applies.} to every quantity with superscript $\Gamma$
\begin{equation}
a^{[\Gamma]}\mapsto a^\nu=\left(a^0,a^1,a^2,a^3\right)\equiv\left(a^{[\gamma^+]},a^{[i\sigma^{1+}\gamma_5]},a^{[i\sigma^{2+}\gamma_5]},a^{[\gamma^+\gamma_5]}\right).
\end{equation}
Using this notation, we obtain 
\begin{equation}
M^{\nu\lambda'\lambda}=(\bar\sigma^\nu)^{\lambda'\lambda}
\end{equation}
with $\bar\sigma^\nu=(\mathds{1},\usigma)$ and $\usigma$ the three Pauli matrices. Since $\bar\sigma^0$ and $\bar\sigma^3$ are diagonal, the correlators $\Phi^{[\gamma^+]}_q(x,\uk_\perp,\uS)$ and $\Phi^{[\gamma^+\gamma_5]}_q(x,\uk_\perp,\uS)$ have a simple probabilistic interpretation. The former gives the density in momentum space of quarks with flavor $q$ irrespective of their polarization, while the latter gives the net density in momentum space of longitudinally polarized quarks with flavor $q$, \emph{i.e.} the density of quarks with positive light-cone helicity \emph{minus} the density of quarks with negative light-cone helicity. On the contrary, the correlators $\Phi^{[i\sigma^{j+}\gamma_5]}_q(x,\uk_\perp,\uS)$ do not have a simple interpretation in the quark light-cone helicity basis. One can however choose to work in another basis of light-cone polarization. The quark creation operators with light-cone polarization parallel or opposite to the generic direction $\us=\left(\sin\theta_s\,\cos\phi_s,\sin\theta_s\,\sin\phi_s,\cos\theta_s\right)$ can be written in terms of quark creation operators with light-cone helicity $\lambda$ as follows
\begin{equation}
\begin{pmatrix}q^\dag_{+\us},&q^\dag_{-\us}\end{pmatrix}=\begin{pmatrix}q^\dag_+,&q^\dag_-\end{pmatrix}u(\theta_s,\phi_s),
\end{equation}
where the $SU(2)$ rotation matrix $u(\theta,\phi)$ is given by
\begin{equation}\label{su2rot}
u(\theta,\phi)=\begin{pmatrix}\cos\tfrac{\theta}{2}\,e^{-i\phi/2}&-\sin\tfrac{\theta}{2}\,e^{-i\phi/2}\\\sin\tfrac{\theta}{2}\,e^{i\phi/2}&\cos\tfrac{\theta}{2}\,e^{i\phi/2}\end{pmatrix}.
\end{equation}
In this way, we see that the correlators $\Phi^{[i\sigma^{1+}\gamma_5]}_q(x,\uk_\perp,\uS)$ and $\Phi^{[i\sigma^{2+}\gamma_5]}_q(x,\uk_\perp,\uS)$ give the net density in momentum space of quarks with flavor $q$ and light-cone polarization in the direction $\boldsymbol e_x$ and $\boldsymbol e_y$, respectively. Clearly, the net density of quarks with generic light-cone polarization $\us$ is given by the correlator $\us\cdot\boldsymbol\Phi_q(x,\uk_\perp,\uS)$.

In the literature, one often represents correlators in terms of helicity amplitudes which treat in a symmetric way both quark and target polarizations
\begin{equation}\label{helicityamplitude}
\Phi^q_{\Lambda'\lambda',\Lambda\lambda}(x,\uk_\perp)=\frac{1}{\mathcal N}\,\langle P,\Lambda'|q^\dag_{\lambda'}(\tilde k)\,q_\lambda(\tilde k)|P,\Lambda\rangle.
\end{equation}
Decomposing the target states $|P,\pm\uS\rangle$ with light-cone polarization parallel or opposite to the generic direction $\uS=\left(\sin\theta_S\,\cos\phi_S,\sin\theta_S\,\sin\phi_S,\cos\theta_S\right)$ in terms of the target light-cone helicity states $|P,\Lambda\rangle$
\begin{equation}
\begin{pmatrix}|P,+\uS\rangle,&|P,-\uS\rangle\end{pmatrix}=\begin{pmatrix}|P,+\rangle,&|P,-\rangle\end{pmatrix}u(\theta_S,\phi_S),
\end{equation}
one obtains that the helicity amplitudes are given by the following combinations of TMDs
\begin{equation}\label{amplTMDs}
\Phi^q_{\Lambda'\lambda',\Lambda\lambda}=\begin{pmatrix}
\tfrac{1}{2}\left(f^q_1+g^q_{1L}\right)&-\tfrac{k_R}{2M}\left(ih_1^{\perp q}-h_{1L}^{\perp q}\right)&\tfrac{k_L}{2M}\left(if_{1T}^{\perp q}+g^q_{1T}\right)&h^q_1\\
\tfrac{k_L}{2M}\left(ih_1^{\perp q}+h_{1L}^{\perp q}\right)&\tfrac{1}{2}\left(f^q_1-g^q_{1L}\right)&\tfrac{k_L^2}{2M^2}\,h_{1T}^{\perp q}&\tfrac{k_L}{2M}\left(if_{1T}^{\perp q}-g^q_{1T}\right)\\
-\tfrac{k_R}{2M}\left(if_{1T}^{\perp q}-g^q_{1T}\right)&\tfrac{k_R^2}{2M^2}\,h_{1T}^{\perp q}&\tfrac{1}{2}\left(f^q_1-g^q_{1L}\right)&-\tfrac{k_R}{2M}\left(ih_1^{\perp q}+h_{1L}^{\perp q}\right)\\
h^q_1&-\tfrac{k_R}{2M}\left(if_{1T}^{\perp q}+g^q_{1T}\right)&\tfrac{k_L}{2M}\left(ih_1^{\perp q}-h_{1L}^{\perp q}\right)&\tfrac{1}{2}\left(f^q_1+g^q_{1L}\right)
\end{pmatrix},
\end{equation}
where $k_{R,L}=k_x\pm ik_y$. The rows entries are $(\Lambda'\lambda')=(++),(+-),(-+),(--)$ and the columns entries are likewise $(\Lambda\lambda)=(++),(+-),(-+),(--)$.

We find actually more convenient to represent the correlator \eqref{corrTMD} in the four-component basis by the tensor $\Phi^{\mu\nu}_q(x,\uk_\perp)$. This tensor is related to helicity amplitudes as follows
\begin{equation}\label{basischange}
\Phi^{\mu\nu}_q=\frac{1}{2}\sum_{\Lambda'\Lambda\lambda'\lambda}(\bar\sigma^\mu)^{\Lambda\Lambda'}(\bar\sigma^\nu)^{\lambda'\lambda}\,\Phi^q_{\Lambda'\lambda',\Lambda\lambda},\qquad\Phi^q_{\Lambda'\lambda',\Lambda\lambda}=\frac{1}{2}\,\Phi^{\mu\nu}_q(\sigma_\mu)_{\Lambda'\Lambda}(\sigma_\nu)_{\lambda\lambda'},
\end{equation}
where $\sigma_\mu=g_{\mu\nu}\sigma^\nu$ with $\sigma^\nu=(\mathds{1},-\usigma)$. The symbols $\bar\sigma^\mu$ and $\sigma_\mu$ satisfy the relations
\begin{equation}\label{basisidentities}
\frac{1}{2}\,(\bar\sigma^\mu)^{\lambda'\lambda}(\sigma_\mu)_{\tau\tau'}=\delta^{\lambda'}_{\tau'}\delta^\lambda_\tau,\qquad\frac{1}{2}\uTr\left[\bar\sigma^\mu\sigma_\nu\right]=\frac{1}{2}\sum_{\lambda'\lambda}(\bar\sigma^\mu)^{\lambda'\lambda}(\sigma_\nu)_{\lambda\lambda'}=\delta^\mu_\nu.
\end{equation}
The tensor correlator is then given by the following combinations of TMDs
\begin{align}
\Phi^{\mu\nu}_q&=\begin{pmatrix}
f^q_1&\frac{k_y}{M}\,h^{\perp q}_1&-\frac{k_x}{M}\,h^{\perp q}_1&0\\
\frac{k_y}{M}\,f^{\perp q}_{1T}&h^q_1+\frac{k^2_x-k^2_y}{2M^2}\,h^{\perp q}_{1T}&\frac{k_xk_y}{M^2}\,h^{\perp q}_{1T}&\frac{k_x}{M}\,g^q_{1T}\\
-\frac{k_x}{M}\,f^{\perp q}_{1T}&\frac{k_xk_y}{M^2}\,h^{\perp q}_{1T}&h^q_1-\frac{k^2_x-k^2_y}{2M^2}\,h^{\perp q}_{1T}&\frac{k_y}{M}\,g^q_{1T}\\
0&\frac{k_x}{M}\,h^{\perp q}_{1L}&\frac{k_y}{M}\,h^{\perp q}_{1L}&g^q_{1L}
\end{pmatrix}\nonumber\\
&=\begin{pmatrix}
f^q_1&\tfrac{k_y}{M}\,h_1^{\perp q}&-\tfrac{k_x}{M}\,h_1^{\perp q}&0\\
\tfrac{k_y}{M}\,f_{1T}^{\perp q}&h_{1T}^{+q}\,\hat k_x^2+h_{1T}^{-q}\,\hat k_y^2&\left(h_{1T}^{+q}-h_{1T}^{-q}\right)\hat k_x\hat k_y&\tfrac{k_x}{M}\,g_{1T}^q\\
-\tfrac{k_x}{M}\,f_{1T}^{\perp q}&\left(h_{1T}^{+q}-h_{1T}^{-q}\right)\hat k_x\hat k_y&h_{1T}^{-q}\,\hat k_x^2+h_{1T}^{+q}\,\hat k_y^2&\tfrac{k_y}{M}\,g_{1T}^q\\
0&\tfrac{k_x}{M}\,h_{1L}^{\perp q}&\tfrac{k_x}{M}\,h_{1L}^{\perp q}&g_{1L}^q
\end{pmatrix},\label{tensor2}
\end{align}
where we introduced the notations $h_{1T}^{\pm q}=h^q_1\pm\tfrac{\uk_\perp^2}{2M^2}\,h_{1T}^{\perp q}$ and $\hat k_i=k_i/k_\perp$. The component $\Phi^{00}_q$ gives the density of quarks in the target irrespective of any polarization, \emph{i.e.} the density of unpolarized quarks in the unpolarized target. The components $\Phi^{0j}_q$ give the net density of quarks with light-cone polarization in the direction $\boldsymbol e_j$ in the unpolarized target, while the components $\Phi^{i0}_q$ give the net density of unpolarized quarks in the target with light-cone polarization in the direction $\boldsymbol e_i$. Finally, the components $\Phi^{ij}_q$ give the net density of quarks with light-cone polarization in the direction $\boldsymbol e_j$ in the target with light-cone polarization in the direction $\boldsymbol e_i$. The density of quarks with definite light-cone polarization in the direction $\us$ inside the target with definite light-cone polarization in the direction $\uS$ is then obviously given by $\Phi_q(x,\uk_\perp,\uS,\us)=\frac{1}{2}\,\bar S_\mu\Phi_q^{\mu\nu}\bar s_\nu$, where we have introduced the four-component vectors $\bar S_\mu=(1,\uS)$ and $\bar s_\nu=(1,\us)$.

\section{Model Relations}
\label{section-3}

In QCD, the eight TMDs are in principle independent. It appeared however in a large panel of low-energy models that relations among some TMDs exist. At twist-two level, there are three flavor-independent relations\footnote{Other expressions can be found in the literature, but are just combinations of the relations \eqref{rel1}-\eqref{rel3}.}, two are linear and one is quadratic in the TMDs
\begin{gather}
g^q_{1L}-\left[h^q_1+\tfrac{\uk_\perp^2}{2M^2}\,h_{1T}^{\perp q}\right]=0,\label{rel1}\\
g^q_{1T}+h_{1L}^{\perp q}=0,\label{rel2}\\
\left(g^q_{1T}\right)^2+2h^q_1\,h_{1T}^{\perp q}=0.\label{rel3}
\end{gather}
A further flavor-dependent relation involves both polarized and unpolarized TMDs
\begin{equation}\label{rel4}
\mathcal D^qf_1^q+g_{1L}^q=2h_1^q,
\end{equation}
where, for a proton target, the flavor factors with $q=u,d$ are given by $\mathcal D^u=\tfrac{2}{3}$ and $\mathcal D^d=-\tfrac{1}{3}$. As discussed in Ref.~\cite{Avakian:2010br}, at variance with the relations \eqref{rel1}-\eqref{rel3}, the flavor dependence in the relation \eqref{rel4} requires specific assumptions for the spin-isospin structure of the nucleon state, like $SU(6)$ spin-flavor symmetry.

A discussion on how general these relations are can be found in Ref.~\cite{Avakian:2010br}. Let us just mention that they were observed in the bag model \cite{Avakian:2010br,Avakian:2008dz}, light-cone constituent quark models \cite{Pasquini:2008ax}, some quark-diquark models \cite{Jakob:1997wg,She:2009jq,Zhu:2011zza}, the covariant parton model \cite{Efremov:2009ze} and more recently in the light-cone version of the chiral quark-soliton model \cite{Lorce:2011dv}. Note however that there also exist models where the relations are not satisfied, like in some versions of the spectator model~\cite{Bacchetta:2008af} and the quark-target model \cite{Meissner:2007rx}. 

As already emphasized, the model relations \eqref{rel1}-\eqref{rel4} are not expected to hold identically in QCD, 
since the TMDs in these relations follow different evolution patterns.
This implies that even if the relations are satisfied at some (low) scale, 
they would not hold anymore for other (higher) scales. 
The interest in these relations is therefore purely phenomenological. 
Experiments provide more and more data on observables related to TMDs, but need inputs from educated models and
 parametrizations for the extraction of these distributions.
  It is therefore particularly interesting to see to what extent the relations \eqref{rel1}-\eqref{rel4} can be useful 
as \emph{approximate} relations, which provide simplified and intuitive notions for the interpretation of the data. Note that some preliminary calculations in lattice QCD give indications that the relation \eqref{rel2} may indeed be approximately satisfied \cite{Musch:2010ka,Hagler:2009mb}.

Using two different approaches, we show in the next sections that the flavor-independent relations \eqref{rel1}-\eqref{rel3} can easily be derived, once the following assumptions are made:
\begin{enumerate}
\item 
the probed quark behaves as if it does not interact directly with the other partons ({\em i.e.} one works within the standard impulse approximation) and
there are no explicit gluons;
\item the quark light-cone and canonical polarizations are related by a rotation with axis orthogonal to both light-cone and $\hat k_\perp$ directions;
\item the target has spherical symmetry in the canonical-spin basis.
\end{enumerate}
From these assumptions, one realizes that the flavor-independent relations have essentially a geometrical origin, as was already suspected in the context of the bag model almost a decade ago \cite{Efremov:2002qh}. 
We note however that the spherical symmetry is a sufficient but not necessary condition for the validity
of the individual flavor-independent relations. 
As discussed in the following section, a subset of relations can be derived using less restrictive conditions, like axial symmetry about a specific direction.
\newline
\noindent
For the flavor-dependent relation \eqref{rel4}, we need a further condition for the spin-flavor dependent part
of the nucleon wave function. Specifically, we require 
\begin{enumerate}[resume]
\item $SU(6)$ spin-flavor symmetry of the wave function.
\end{enumerate}
As shown in Sec.~\ref{models}, it is found that all the models satisfying the relations also satisfy the above conditions.
We are not aware of any model calculation which satisfies some
or all the three flavor-independent relations and at the same time breaks at least one of the 
conditions 1-3. However, this is
 not  a priori excluded.

\section{Amplitude Approach}

The first derivation of the TMD relations stays at the level of the amplitudes. As we have seen, the TMDs can be expressed in simple terms using light-cone polarization. On the other hand, rotational symmetry is easier to handle in terms of canonical polarization, which is the natural one in the instant form. We therefore write the TMDs in the canonical-spin basis, and then impose spherical symmetry. But before that, we need to know how to connect light-cone helicity to canonical spin.

\subsection{Connection between Light-Cone Helicity and Canonical Spin}

Relating in general light-cone helicity with canonical spin is usually quite complicated, as the dynamics is involved. 
Fortunately, the common approach in quark models is to assume that the target can be described by quarks without mutual interactions. 
In this case the connection simply reduces to a rotation 
in polarization space with axis orthogonal to both $\uk_\perp$ and $\boldsymbol e_z$ directions. 
The quark creation operator with canonical spin $\sigma$ can then be written in terms of quark creation operators with light-cone helicity $\lambda$ as follows
\begin{equation}\label{genMelosh}
q^\dag_\sigma=\sum_\lambda D^{(1/2)*}_{\sigma\lambda}\,q^\dag_\lambda\qquad\text{with}\qquad D^{(1/2)*}_{\sigma\lambda}=\begin{pmatrix}\cos\tfrac{\theta}{2}&-\hat k_R\,\sin\tfrac{\theta}{2}\\\hat k_L\,\sin\tfrac{\theta}{2}&\cos\tfrac{\theta}{2}\end{pmatrix}.
\end{equation}
Note that the rotation does not depend on the quark flavor. The angle $\theta$ between light-cone and canonical polarizations is usually a complicated function of the quark momentum $k$ and is specific to each model. It contains part of the model dynamics. The only general property is that $\theta\to 0$ as $k_\perp\to 0$. 
Due to our choice of reference frame where the target has no transverse momentum,
the light-cone helicity and canonical spin of the target can be identified, at variance 
with  the quark polarizations.

\subsection{TMDs in Canonical-Spin Basis}

The four-component notation introduced in sect.~\ref{sect:II-B} is very convenient for discussing the rotation between canonical spin and light-cone helicity at the amplitude level. Since the light-cone helicity and canonical spin of the target can be identified in our choice of reference frame, we expect the canonical tensor correlator $\Phi^{\mu\nu}_{Cq}$ to be related to the light-cone one in Eq.~\eqref{tensor2} as follows
\begin{equation}\label{expectation}
\Phi^{\mu\nu}_{Cq}=\Phi_q^{\mu\rho}\,O_\rho^{\phantom{\rho}\nu},
\end{equation}
with $O$ some orthogonal matrix $O^TO=\mathds 1$ representing the rotation at the amplitude level. From Eqs.~\eqref{helicityamplitude}, \eqref{basischange}, \eqref{basisidentities} and \eqref{genMelosh} we find that the orthogonal matrix is given by
\begin{align}
O_\rho^{\phantom{\rho}\nu}&=\frac{1}{2}\uTr\left[D^{(1/2)}\sigma_\rho D^{(1/2)\dag}\bar\sigma^\nu\right]\nonumber\\
&=\frac{1}{2}\sum_{\sigma'\sigma\lambda'\lambda}D^{(1/2)}_{\sigma\lambda}\left(\sigma_\rho\right)_{\lambda\lambda'}\,D^{(1/2)*}_{\sigma'\lambda'}\left(\bar\sigma^\nu\right)^{\sigma'\sigma}\nonumber\\
&=\begin{pmatrix}
1&0&0&0\\
0&\hat k_y^2+\hat k_x^2\,\cos\theta&-\hat k_x\hat k_y\left(1-\cos\theta\right)&-\hat k_x\,\sin\theta\\
0&-\hat k_x\hat k_y\left(1-\cos\theta\right)&\hat k_x^2+\hat k_y^2\,\cos\theta&-\hat k_y\,\sin\theta\\
0&\hat k_x\,\sin\theta&\hat k_y\,\sin\theta&\cos\theta
\end{pmatrix}.\label{ortho}
\end{align}
The canonical tensor correlator then takes the form
\begin{equation}\label{Ctensor}
\Phi^{\mu\nu}_{Cq}=\begin{pmatrix}
f^q_1&\tfrac{k_y}{M}\,h_1^{\perp q}&-\tfrac{k_x}{M}\,h_1^{\perp q}&0\\
\tfrac{k_y}{M}\,f_{1T}^{\perp q}&\mathfrak h_{1T}^{+q}\,\hat k_x^2+h_{1T}^{-q}\,\hat k_y^2&\left(\mathfrak h_{1T}^{+q}-h_{1T}^{-q}\right)\hat k_x\hat k_y&\tfrac{k_x}{M}\,\mathfrak g_{1T}^q\\
-\tfrac{k_x}{M}\,f_{1T}^{\perp q}&\left(\mathfrak h_{1T}^{+q}-h_{1T}^{-q}\right)\hat k_x\hat k_y&h_{1T}^{-q}\,\hat k_x^2+\mathfrak h_{1T}^{+q}\,\hat k_y^2&\tfrac{k_y}{M}\,\mathfrak g_{1T}^q\\
0&\tfrac{k_x}{M}\,\mathfrak h_{1L}^{\perp q}&\tfrac{k_x}{M}\,\mathfrak h_{1L}^{\perp q}&\mathfrak g_{1L}^q
\end{pmatrix},
\end{equation}
where we introduced the notations
\begin{align}
\begin{pmatrix}
\mathfrak g^q_{1L}\\ \tfrac{k_\perp}{M}\,\mathfrak h_{1L}^{\perp q}
\end{pmatrix}
&=\begin{pmatrix}\cos\theta&-\sin\theta\\\sin\theta&\cos\theta\end{pmatrix}\begin{pmatrix}
g^q_{1L}\\ \tfrac{k_\perp}{M}\,h_{1L}^{\perp q}
\end{pmatrix},\label{rot1}\\
\begin{pmatrix}
\tfrac{k_\perp}{M}\,\mathfrak g^q_{1T}\\ \mathfrak h_{1T}^{+q}
\end{pmatrix}
&=\begin{pmatrix}\cos\theta&-\sin\theta\\\sin\theta&\cos\theta\end{pmatrix}\begin{pmatrix}
\tfrac{k_\perp}{M}\,g^q_{1T}\\ h_{1T}^{+q}
\end{pmatrix}.\label{rot2}
\end{align}
Comparing Eq.~\eqref{Ctensor} with Eq.~\eqref{tensor2}, we observe that the multipole structure is conserved 
under the rotation~\eqref{expectation}. 
The rotation from light-cone to canonical polarizations affects only some of the multipole magnitudes, see Eqs.~\eqref{rot1} and \eqref{rot2}.

\begin{figure}[ht!]
\vspace{1 truecm}
\begin{center}
\epsfig{file=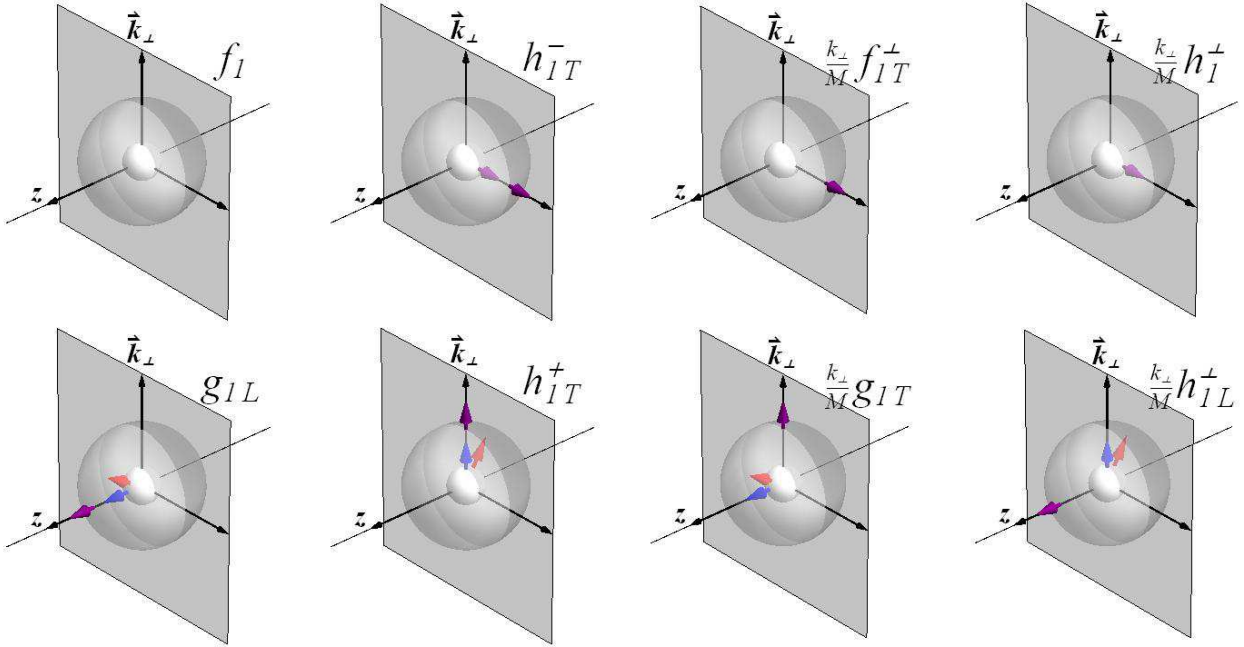,width=\columnwidth}
\end{center}
\caption{\footnotesize{Connection between the TMDs and the amplitudes in the 
net-polarization basis. The $z$-axis corresponds to the light-cone direction. The $y$-axis in the transverse plane has been chosen parallel to the active quark transverse momentum $\uk_\perp$. The arrows attached to the inner and outer spheres represent the net polarizations of the active quark and target, respectively, and absence of arrows represents the unpolarized case. Since we work in a frame where the target has no transverse momentum, there is no difference between target light-cone and canonical polarizations (purple outer arrows). TMDs in the light-cone basis are related to matrix elements where the quark net polarization is along the axes (blue inner arrows), while in the canonical-spin basis the component of quark net polarization in the $(y,z)$-plane is tilted with respect to the axes (red inner arrows), see text.}}
\label{fig1}
\end{figure}

Note that the orientation of the axes in the transverse plane has been fixed arbitrarily. There is however a privileged direction given by the active quark transverse momentum $\uk_\perp$. Choosing the orientation of transverse axes so that either $\uk_\perp=k_\perp\,\boldsymbol e_x$ or $\uk_\perp=k_\perp\,\boldsymbol e_y$ simplifies the transformation, as it eliminates the cumbersome factors $\hat k_x$ and $\hat k_y$ in Eqs.~(\ref{ortho}) and (\ref{Ctensor}).
Choosing \emph{e.g} the second option, the orthogonal matrix of Eq.~\eqref{ortho} reduces to
\begin{equation}
O_\rho^{\phantom{\rho}\nu}=\begin{pmatrix}
\phantom{0}1\phantom{0}&0&0&0\\
0&\phantom{0}1\phantom{0}&0&0\\
0&0&\cos\theta&-\sin\theta\\
0&0&\sin\theta&\cos\theta
\end{pmatrix},
\end{equation}
and the light-cone and canonical tensor correlators take the following simpler forms
\begin{align}
\Phi^{\mu\nu}_q&=\begin{pmatrix}
f^q_1&\frac{k_\perp}{M}\,h^{\perp q}_1&0&0\\
\frac{k_\perp}{M}\,f^{\perp q}_{1T}&h_{1T}^{-q}&0&0\\
0&0&h_{1T}^{+q}&\frac{k_\perp}{M}\,g^q_{1T}\\
0&0&\frac{k_\perp}{M}\,h^{\perp q}_{1L}&g^q_{1L}
\end{pmatrix},\\
\Phi^{\mu\nu}_{Cq}&=\begin{pmatrix}\label{Ctensorred}
f^q_1&\frac{k_\perp}{M}\,h^{\perp q}_1&0&0\\
\frac{k_\perp}{M}\,f^{\perp q}_{1T}&h_{1T}^{-q}&0&0\\
0&0&\mathfrak h_{1T}^{+q}&\frac{k_\perp}{M}\,\mathfrak g_{1T}^q\\
0&0&\frac{k_\perp}{M}\,\mathfrak h^{\perp q}_{1L}&\mathfrak g^q_{1L}
\end{pmatrix}.
\end{align}
Fig.~\ref{fig1} shows graphically the connection between the TMDs and the matrix elements in the four-component (or net-polarization) basis. 

Playing a little bit with Eqs.~\eqref{rot1} and \eqref{rot2}, we find
\begin{equation}\label{lineq}
\begin{pmatrix}
\tfrac{k_\perp}{M}\left(g^q_{1T}+h_{1L}^{\perp q}\right)\\ g^q_{1L}-h_{1T}^{+q}
\end{pmatrix}
=\begin{pmatrix}\cos\theta&-\sin\theta\\\sin\theta&\cos\theta\end{pmatrix}\begin{pmatrix}
\tfrac{k_\perp}{M}\left(\mathfrak g^q_{1T}+\mathfrak h_{1L}^{\perp q}\right)\\ \mathfrak g^q_{1L}-\mathfrak h_{1T}^{+q}
\end{pmatrix},
\end{equation}
and three expressions invariant under the rotation \eqref{expectation}
\begin{align}
\left(\tfrac{k_\perp}{M}\,g^q_{1T}\right)^2+\left(h_{1T}^{+q}\right)^2&=\left(\tfrac{k_\perp}{M}\,\mathfrak g_{1T}^q\right)^2+\left(\mathfrak h_{1T}^{+q}\right)^2,\label{quadeq1}\\
\left(g^q_{1L}\right)^2+\left(\tfrac{k_\perp}{M}\,h_{1L}^{\perp q}\right)^2&=\left(\mathfrak g_{1L}^q\right)^2+\left(\tfrac{k_\perp}{M}\,\mathfrak h_{1L}^{\perp q}\right)^2,\label{quadeq2}\\
g^q_{1L}\,g^q_{1T}+h_{1L}^{\perp q}\,h_{1T}^{+q}&=\mathfrak g_{1L}^q\,\mathfrak g_{1T}^q+\mathfrak h_{1L}^{\perp q}\,\mathfrak h_{1T}^{+q}\label{quadeq3}.
\end{align}
These three invariant expressions have a simple geometric interpretation. The three-component vector $\sum_iS^i\Phi_q^{ij}\equiv\pi^{qj}_{\uS}$ represents the net light-cone polarization of a quark with three-momentum $(xP^+,\uk_\perp)$ and flavor $q$ in a target with net polarization in the $\uS$-direction. From Eq.~\eqref{expectation}, we see that the vector $\upi^q_{C\uS}$ representing the net canonical polarization of the quark is obtained by a rotation of $\upi^q_{\uS}$ 
\begin{equation}
\pi^{qj}_{C\uS}=\sum_{k}\pi^{qk}_{\uS}\,O^{kj}.
\end{equation}
It follows automatically that $\upi^q_{\uS}\cdot\upi^q_{\uS'}$ is invariant under the rotation \eqref{expectation}
\begin{equation}\label{invariance}
\upi^q_{\uS}\cdot\upi^q_{\uS'}=\upi^q_{C\uS}\cdot\upi^q_{C\uS'}.
\end{equation}
Expressions \eqref{quadeq1} and \eqref{quadeq2} are obtained from \eqref{invariance} for the cases $\uS=\uS'=\boldsymbol e_\perp$ and $\uS=\uS'=\boldsymbol e_z$, respectively. They just express the fact  that the magnitude of quark net polarization is the same in the light-cone helicity and canonical-spin bases. Expression \eqref{quadeq3} is obtained for the case $\uS=\boldsymbol e_\perp$ and $\uS'=\boldsymbol e_z$. All the remaining cases do not lead to new independent expressions.

\subsection{Spherical Symmetry}

\begin{figure}[t!]
\begin{center}
\epsfig{file=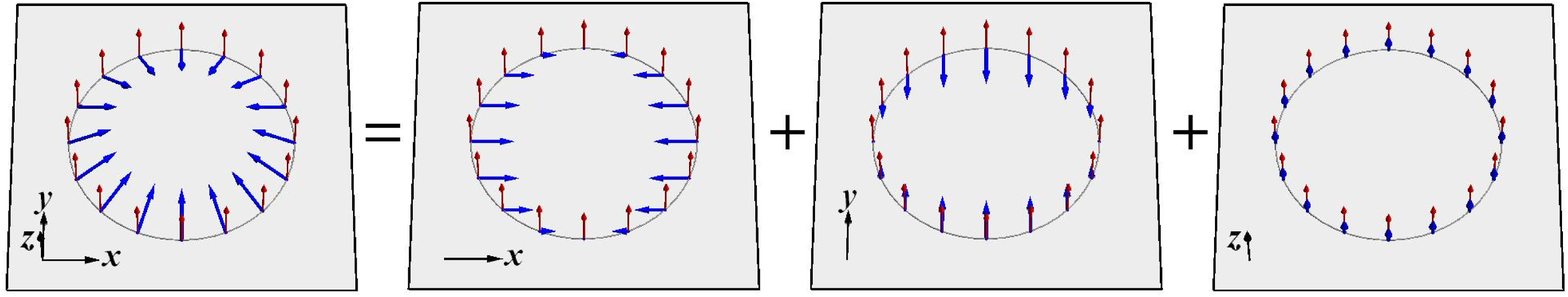,width=\columnwidth}
\end{center}
\caption{\footnotesize{Net light-cone polarization (thick blue arrows) associated to a quark with net longitudinal canonical polarization (thin red arrows), and its vector decomposition along the three axes, for fixed $x$ and $k_\perp$ but arbitrary direction $\hat k_\perp$. The $x$- and $y$-components are pure dipoles, while the $z$-component is a pure monopole.}}
\label{fig2}
\end{figure}

\begin{figure}[t!]
\begin{center}
\epsfig{file=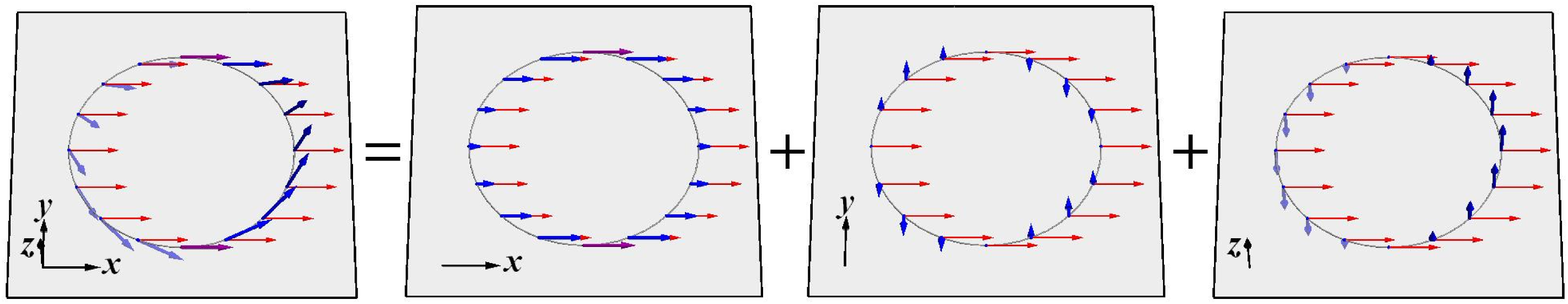,width=\columnwidth}\vspace{.75cm}
\epsfig{file=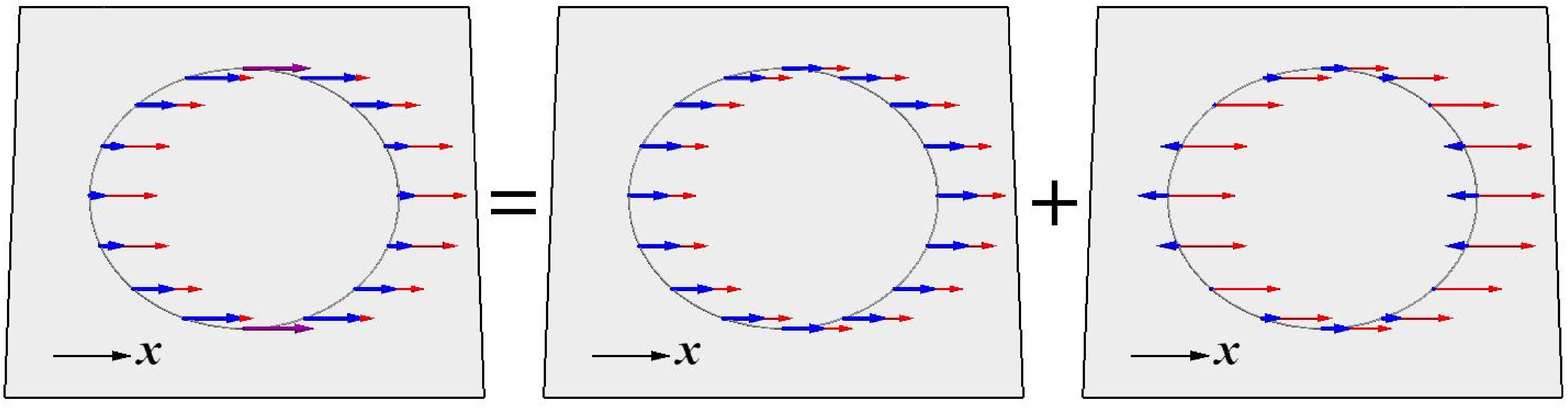,width=.75\columnwidth}
\end{center}
\caption{\footnotesize{In the first line is shown the net light-cone polarization (thick blue arrows) associated to a quark with net canonical polarization in the $x$-direction (thin red arrows), and its vector decomposition along the three axes, for fixed $x$ and $k_\perp$ but arbitrary direction $\hat k_\perp$.  The $y$-component is a pure quadrupole, and the $z$-component is a pure dipole.
The  $x$-component is the sum of a monopole and a quadrupole, as 
illustrated in the second line.}}
\label{fig3}
\end{figure}

We are now ready to discuss the implications of spherical symmetry in the canonical-spin basis. Spherical symmetry means that the canonical tensor correlator has to be invariant $O_R^T\Phi_{Cq}O_R=\Phi_{Cq}$ under any spatial rotation $O_R=\left(\begin{smallmatrix}1&0\\0&R\end{smallmatrix}\right)$ with $R$ the ordinary $3\times 3$ rotation matrix. It is equivalent to the statement that the tensor correlator has to commute with all the elements of the rotation group $\Phi_{Cq}O_R=O_R\Phi_{Cq}$. As a result of Schur's lemma, the canonical tensor correlator must have the following structure
\begin{equation}
\Phi^{\mu\nu}_{Cq}=\begin{pmatrix}A^q&0&0&0\\0&B^q&0&0\\0&0&B^q&0\\0&0&0&B^q\end{pmatrix}.
\end{equation}
Comparing this with Eqs.~\eqref{Ctensor} or \eqref{Ctensorred}, we conclude that spherical symmetry implies
\begin{gather}
f^q_1=A^q,\label{constraint0}\\
\mathfrak g_{1L}^q=\mathfrak h_{1T}^{+q}=h_{1T}^{-q}=B^q,\label{constraint1}\\
\mathfrak g_{1T}^q=\mathfrak h_{1L}^{\perp q}=f_{1T}^{\perp q}=h_1^{\perp q}=0.\label{constraint2}
\end{gather}
Clearly, only the monopole structures in the canonical-spin basis are allowed to survive. In particular, the Sivers and Boer-Mulders functions $f_{1T}^{\perp q}$ and $h_1^{\perp q}$ have to vanish identically. However, as one can see from Figs.~\ref{fig2} and \ref{fig3}, the monopole structures in the canonical-spin basis generate higher multipole structures in the light-cone helicity basis. It follows that spherical symmetry imposes some relations among the multipole structures in the light-cone helicity basis, and therefore among the TMDs.

Inserting the constraints \eqref{constraint1} and \eqref{constraint2} into Eq.~\eqref{lineq}, we automatically obtain the linear relations\footnote{One can also easily understand why spherical symmetry implies $g_{1L}=+h_{1T}^+$ and $g_{1T}=-h_{1L}^\perp$ directly from Fig.~\ref{fig1}. If one performs a $\tfrac{\pi}{2}$-rotation about the $x$-axis on the quark and target polarizations in the representation of $g_{1L}$, one gets the representation of $h_{1T}^+$. If one performs the same transformation on the representation of $\frac{k_\perp}{M}\,g_{1T}$, one gets the representation of $\frac{k_\perp}{M}\,h_{1L}^\perp$ but with one of the net light-cone polarizations in the opposite direction, explaining minus sign.} \eqref{rel1} and \eqref{rel2}
\begin{equation}
\begin{pmatrix}
\tfrac{k_\perp}{M}\left(g^q_{1T}+h_{1L}^{\perp q}\right)\\ g^q_{1L}-h_{1T}^{+q}
\end{pmatrix}
=\begin{pmatrix}
0\\ 0
\end{pmatrix}.
\end{equation}
Using now the constraints from spherical symmetry in Eq.~\eqref{quadeq1}, we obtain the quadratic relation~\eqref{rel3}
\begin{equation}\label{rel3new}
0=\left(\tfrac{k_\perp}{M}\,g^q_{1T}\right)^2+\left(h_{1T}^{+q}\right)^2-\left(h_{1T}^{-q}\right)^2=\tfrac{\uk_\perp^2}{M^2}\left[\left(g^q_{1T}\right)^2+2h^q_1\,h_{1T}^{\perp q}\right].
\end{equation}
The linear relations \eqref{rel1} and \eqref{rel2} being satisfied, the Eqs.~\eqref{quadeq2} and \eqref{quadeq3} do not lead to independent quadratic relations.

We have seen that spherical symmetry is a \emph{sufficient} condition\footnote{From Eqs.~\eqref{lineq} and \eqref{quadeq1}, one can see that the \emph{minimal} conditions are actually
\begin{gather*}
\mathfrak g_{1L}^q-\mathfrak h_{1T}^{+q}=0,\\
\mathfrak g_{1T}^q+\mathfrak h_{1L}^{\perp q}=0,\\
\left(\tfrac{k_\perp}{M}\,\mathfrak g_{1T}^q\right)^2+\left(\mathfrak h_{1T}^{+q}\right)^2-\left(h_{1T}^{-q}\right)^2=0.
\end{gather*}
They are indeed fulfilled by spherical symmetry, see Eqs.~\eqref{constraint1} and \eqref{constraint2}.} to obtain all three flavor-independent relations. At the same time, it tells us that the Sivers and Boer-Mulders distributions have to vanish identically. Restricting ourselves to axial symmetries, we find that some of the relations can already be obtained. For example, axial symmetry about $\boldsymbol e_z$ alone implies $f_{1T}^\perp=h_1^\perp=0$, the quadratic relation \eqref{rel3} and 
\begin{equation}\label{newquadratic}
g^q_{1L}\,g^q_{1T}+h_{1L}^{\perp q} \,h_{1T}^{+q}=0.
\end{equation}
Axial symmetry about $\hat k_\perp\times\boldsymbol e_z$ implies the two linear relations \eqref{rel1} and \eqref{rel2}. The relation \eqref{newquadratic} is naturally also satisfied but is not independent.

\section{Wave-Function Approach}
\label{section-5}

Many quark models are based on a wave-function approach. We therefore translate here the derivation of the previous section in the language of 3Q wave functions. The advantage is that we can then also discuss the additional $SU(6)$ spin-flavor symmetry needed for the flavor-dependent relation \eqref{rel4}. For convenience, we omit all the color indices in the following expressions.

\subsection{Overlap Representation of the TMDs on the Light Cone}

Restricting ourselves to the 3Q Fock sector, the target state with definite four-momentum $P=[P^+,\tfrac{M^2}{2P^+},\uzero_\perp]$ and light-cone helicity $\Lambda$ can be written as follows
\begin{equation}\label{LCWF}
|P,\Lambda\rangle=\sum_{\lambda_1\lambda_2\lambda_3}\sum_{q_1q_2q_3}\int[\ud x]_3\,[\ud^2k_\perp]_3\,\psi^{\Lambda;q_1q_2q_3}_{\lambda_1\lambda_2\lambda_3}(\tilde k_1,\tilde k_2,\tilde k_3)\,|\{\lambda_i,q_i,\tilde k_i\}\rangle,
\end{equation}
where $\psi^{\Lambda;q_1q_2q_3}_{\lambda_1\lambda_2\lambda_3}(\tilde k_1,\tilde k_2,\tilde k_3)$ is the three-quark light-cone wave function (3Q LCWF) with $\lambda_i$, $q_i$ and $\tilde k_i$ referring to the light-cone helicity, flavor and light-cone momentum of quark $i$, respectively. The total orbital angular momentum of a given component $\psi^\Lambda_{\lambda_1\lambda_2\lambda_3}$ is given by the expression $\ell_z=\Lambda-\lambda_1-\lambda_2-\lambda_3$ with $\Lambda,\lambda_i=\pm\tfrac{1}{2}$. The integration measures in Eq.~\eqref{LCWF} are defined as
\begin{equation}
\begin{split}
[\ud x]_3&\equiv\left[\prod_{i=1}^3\ud x_i\right]\delta\!\!\left(1-\sum_{i=1}^3x_i\right),\\
[\ud^2k_\perp]_3&\equiv\left[\prod_{i=1}^3\frac{\ud^2k_{i\perp}}{2(2\pi)^3}\right]2(2\pi)^3\,\delta^{(2)}\!\!\left(\sum_{i=1}^3\uk_{i\perp}\right).
\end{split}
\end{equation}
Choosing to label the active quark with $i=1$, the TMDs can be obtained by the following overlaps\footnote{In the 3Q approach, the spectator system consists of two quarks. It is straightforward to generalize the expression for helicity amplitudes to any kind of spectator system, as the latter is integrated out.} of 3Q LCWFs
\begin{subequations}
\begin{align}
f^q_1&=\int\ud[23]\sum_{\lambda_2\lambda_3}\sum_{q_2q_3}\left[|\psi^{+;qq_2q_3}_{+\lambda_2\lambda_3}|^2+|\psi^{+;qq_2q_3}_{-\lambda_2\lambda_3}|^2\right],\\
g^q_{1L}&=\int\ud[23]\sum_{\lambda_2\lambda_3}\sum_{q_2q_3}\left[|\psi^{+;qq_2q_3}_{+\lambda_2\lambda_3}|^2-|\psi^{+;qq_2q_3}_{-\lambda_2\lambda_3}|^2\right],\\
h^q_1&=\int\ud[23]\sum_{\lambda_2\lambda_3}\sum_{q_2q_3}\left(\psi^{+;qq_2q_3}_{+\lambda_2\lambda_3}\right)^*\psi^{-;qq_2q_3}_{-\lambda_2\lambda_3},\\
\tfrac{k_\perp}{M}\,f^{\perp q}_{1T}&=\int\ud[23]\sum_{\lambda_2\lambda_3}\sum_{q_2q_3}2\,\Im m\left[\hat k_R\left(\psi^{+;qq_2q_3}_{+\lambda_2\lambda_3}\right)^*\psi^{-;qq_2q_3}_{+\lambda_2\lambda_3}\right],\\
\tfrac{k_\perp}{M}\,g^q_{1T}&=\int\ud[23]\sum_{\lambda_2\lambda_3}\sum_{q_2q_3}2\,\Re e\left[\hat k_R\left(\psi^{+;qq_2q_3}_{+\lambda_2\lambda_3}\right)^*\psi^{-;qq_2q_3}_{+\lambda_2\lambda_3}\right],\\
\tfrac{k_\perp}{M}\,h^{\perp q}_1&=\int\ud[23]\sum_{\lambda_2\lambda_3}\sum_{q_2q_3}2\,\Im m\left[\hat k_R\left(\psi^{+;qq_2q_3}_{-\lambda_2\lambda_3}\right)^*\psi^{+;qq_2q_3}_{+\lambda_2\lambda_3}\right],\\
\tfrac{k_\perp}{M}\,h^{\perp q}_{1L}&=\int\ud[23]\sum_{\lambda_2\lambda_3}\sum_{q_2q_3}2\,\Re e\left[\hat k_R\left(\psi^{+;qq_2q_3}_{-\lambda_2\lambda_3}\right)^*\psi^{+;qq_2q_3}_{+\lambda_2\lambda_3}\right],\\
\tfrac{\uk_\perp^2}{2M^2}\,h^{\perp q}_{1T}&=\int\ud[23]\sum_{\lambda_2\lambda_3}\sum_{q_2q_3}\hat k^2_R\left(\psi^{+;qq_2q_3}_{-\lambda_2\lambda_3}\right)^*\psi^{-;qq_2q_3}_{+\lambda_2\lambda_3},
\end{align}
\end{subequations}
where we used the notation
\begin{equation}
\ud[23]=[\ud x]_3\,[\ud^2k_\perp]_3\,3\,\delta(x-x_1)\,\delta^{(2)}(\uk_\perp-\uk_{1\perp}).
\end{equation}
Clearly, the TMDs associated to the monopole structures ($f^q_1,g^q_{1L},h^q_1$) are represented by overlaps with no global change of orbital angular momentum $\Delta\ell_z=0$, the ones associated to the dipole structures ($f_{1T}^{\perp q},g^q_{1T},h_1^{\perp q},h_{1L}^{\perp q}$) involve a change by one unit of orbital angular momentum $|\Delta\ell_z|=1$ and the one associated to the quadrupole structure ($h_{1T}^{\perp q}$) involves a change by two units of orbital angular momentum $|\Delta\ell_z|=2$.

\subsection{Overlap Representation of the TMDs in the Canonical-Spin Basis}

Most of the quark models being originally formulated in the instant form, it is more natural to work in the canonical-spin basis instead of the light-cone helicity basis. Since we considered a frame where the target has no transverse momentum, there is no difference between target light-cone and canonical polarizations. Assuming that the quark light-cone helicity and canonical spin are connected by the rotation in Eq.~\eqref{genMelosh}, 
the components of the LCWF
in the canonical-spin basis $\psi^\Lambda_{\sigma_1\sigma_2\sigma_3}$ (with $\sigma_i=\uparrow,\downarrow$)
and in the light-cone helicity basis
$\psi^\Lambda_{\lambda_1\lambda_2\lambda_3}$ (with $\lambda_i=\pm$)
are related as follows\footnote{Note that $\psi^\Lambda_{\sigma_1\sigma_2\sigma_3}$ cannot be identified in general with the usual rest-frame wave function $ \Psi^\Lambda_{\sigma_1\sigma_2\sigma_3}$. They have the same spin 
structure, but not  the same momentum dependence.}
\begin{equation}\label{connection}
\psi^\Lambda_{\lambda_1\lambda_2\lambda_3}=\sum_{\sigma_1\sigma_2\sigma_3}\psi^\Lambda_{\sigma_1\sigma_2\sigma_3}\,D^{(1/2)*}_{\sigma_1\lambda_1}\,D^{(1/2)*}_{\sigma_2\lambda_2}\,D^{(1/2)*}_{\sigma_3\lambda_3}.
\end{equation}
The correspondence between the components in the two polarization bases is given in a more explicit form in App. \ref{table}. Using $D^{(1/2)\dag}D^{(1/2)}=\mathds 1$, we find the explicit overlap representations in canonical-spin basis
\begin{subequations}
\begin{align}
f^q_1&=\int\ud[23]\sum_{\sigma_2\sigma_3}\sum_{q_2q_3}\left[|\psi^{\uparrow;qq_2q_3}_{\uparrow\sigma_2\sigma_3}|^2+|\psi^{\uparrow;qq_2q_3}_{\downarrow\sigma_2\sigma_3}|^2\right],\\
\mathfrak g^q_{1L}&=\int\ud[23]\sum_{\sigma_2\sigma_3}\sum_{q_2q_3}\left[|\psi^{\uparrow;qq_2q_3}_{\uparrow\sigma_2\sigma_3}|^2-|\psi^{\uparrow;qq_2q_3}_{\downarrow\sigma_2\sigma_3}|^2\right],\\
\mathfrak h_1^q&=\int\ud[23]\sum_{\sigma_2\sigma_3}\sum_{q_2q_3}\left(\psi^{\uparrow;qq_2q_3}_{\uparrow\sigma_2\sigma_3}\right)^*\psi^{\downarrow;qq_2q_3}_{\downarrow\sigma_2\sigma_3},\\
\tfrac{k_\perp}{M}\,f^{\perp q}_{1T}&=\int\ud[23]\sum_{\sigma_2\sigma_3}\sum_{q_2q_3}2\,\Im m\left[\hat k_R\left(\psi^{\uparrow;qq_2q_3}_{\uparrow\sigma_2\sigma_3}\right)^*\psi^{\downarrow;qq_2q_3}_{\uparrow\sigma_2\sigma_3}\right],\\
\tfrac{k_\perp}{M}\,\mathfrak g^q_{1T}&=\int\ud[23]\sum_{\sigma_2\sigma_3}\sum_{q_2q_3}2\,\Re e\left[\hat k_R\left(\psi^{\uparrow;qq_2q_3}_{\uparrow\sigma_2\sigma_3}\right)^*\psi^{\downarrow;qq_2q_3}_{\uparrow\sigma_2\sigma_3}\right],\\
\tfrac{k_\perp}{M}\,h^\perp_1&=\int\ud[23]\sum_{\sigma_2\sigma_3}\sum_{q_2q_3}2\,\Im m\left[\hat k_R\left(\psi^{\uparrow;qq_2q_3}_{\downarrow\sigma_2\sigma_3}\right)^*\psi^{\uparrow;qq_2q_3}_{\uparrow\sigma_2\sigma_3}\right],\\
\tfrac{k_\perp}{M}\,\mathfrak h^{\perp q}_{1L}&=\int\ud[23]\sum_{\sigma_2\sigma_3}\sum_{q_2q_3}2\,\Re e\left[\hat k_R\left(\psi^{\uparrow;qq_2q_3}_{\downarrow\sigma_2\sigma_3}\right)^*\psi^{\uparrow;qq_2q_3}_{\uparrow\sigma_2\sigma_3}\right],\\
\tfrac{\uk_\perp^2}{2M^2}\,\mathfrak h^{\perp q}_{1T}&=\int\ud[23]\sum_{\sigma_2\sigma_3}\sum_{q_2q_3}\hat k^2_R\left(\psi^{\uparrow;qq_2q_3}_{\downarrow\sigma_2\sigma_3}\right)^*\psi^{\downarrow;qq_2q_3}_{\uparrow\sigma_2\sigma_3},
\end{align}
\end{subequations}
with $\mathfrak h_{1T}^{+q}=\mathfrak h^q_1+\tfrac{\uk_\perp^2}{2M^2}\,\mathfrak h_{1T}^{\perp q}$ and $h_{1T}^{-q}=\mathfrak h^q_1-\tfrac{\uk_\perp^2}{2M^2}\,\mathfrak h_{1T}^{\perp q}$. The functions $\mathfrak g^q_{1L},\mathfrak g^q_{1T},\mathfrak h^{\perp q}_{1L},\mathfrak h^{+ q}_{1T}$ are again related to the TMDs $g^q_{1L},g^q_{1T},h^{\perp q}_{1L},h^{+ q}_{1T}$ according to Eqs.~\eqref{rot1} and \eqref{rot2}.

\subsection{Spherical Symmetry}

We now discuss how spherical symmetry restricts the form of the wave function in the canonical-spin basis. Spherical symmetry requires the wave function to be invariant under any rotation, i.e.
\begin{equation}
\sum_{\Lambda'\sigma'_1\sigma'_2\sigma'_3}\left[u(\theta,\phi)\right]_{\sigma_1\sigma'_1}\left[u(\theta,\phi)\right]_{\sigma_2\sigma'_2}\left[u(\theta,\phi)\right]_{\sigma_3\sigma'_3}\left[u(\theta,\phi)\right]^*_{\Lambda\Lambda'}\psi^{\Lambda'}_{\sigma'_1\sigma'_2\sigma'_3}=\psi^\Lambda_{\sigma_1\sigma_2\sigma_3},
\end{equation}
with the $SU(2)$ rotation matrix $u(\theta,\phi)$ given by Eq.~\eqref{su2rot}. In particular, invariance under a $(\pi,0)$-rotation leads to
\begin{equation}\label{rotinv1}
\psi^{-\Lambda}_{-\sigma_1-\sigma_2-\sigma_3}=(-1)^{\Lambda+\sigma_1+\sigma_2+\sigma_3}\,\psi^\Lambda_{\sigma_1\sigma_2\sigma_3},
\end{equation}
while invariance under $(0,\phi)$-rotations implies that all components with $\ell_z\neq 0$ have to vanish
\begin{equation}\label{rotinv2}
\psi^\uparrow_{\uparrow\uparrow\uparrow}=\psi^\uparrow_{\downarrow\downarrow\uparrow}=\psi^\uparrow_{\downarrow\uparrow\downarrow}=\psi^\uparrow_{\uparrow\downarrow\downarrow}=\psi^\uparrow_{\downarrow\downarrow\downarrow}=0.
\end{equation}
Taking into account the constraints \eqref{rotinv1} and \eqref{rotinv2} in an arbitrary $(\theta,\phi)$-rotation, one finally gets\footnote{Note that spherical symmetry neither restricts the number of non-zero components of the wave function in the light-cone helicity basis nor relates them in a simple way, see Table~\ref{3QLCWF} in App.~\ref{table}.}
\begin{equation}\label{rotinv3}
\psi^\uparrow_{\uparrow\uparrow\downarrow}+\psi^\uparrow_{\uparrow\downarrow\uparrow}+\psi^\uparrow_{\downarrow\uparrow\uparrow}=0.
\end{equation}

Again, spherical symmetry implies that the TMDs are either identically zero or proportional to the unpolarized and polarized amplitudes $A^q$ and $B^q$,
\begin{subequations}\label{sphericalTMDs}
\begin{align}
f^q_1&=A^q,\\
g^q_{1L}&=\cos\theta\,B^q,\\
h^q_1&=\frac{\cos\theta+1}{2}\,B^q,\\
\tfrac{k_\perp}{M}\,f^{\perp q}_{1T}&=0,\\
\tfrac{k_\perp}{M}\,g^q_{1T}&=\sin\theta\,B^q,\\
\tfrac{k_\perp}{M}\,h^{\perp q}_1&=0,\\
\tfrac{k_\perp}{M}\,h^{\perp q}_{1L}&=-\sin\theta\,B^q,\\
\tfrac{\uk_\perp^2}{2M^2}\,h^{\perp q}_{1T}&=\frac{\cos\theta-1}{2}\,B^q,
\end{align}
\end{subequations}
with $A^q$ and $B^q$ given by the following overlaps 
\begin{align}
A^q&=\int\ud[23]\sum_{q_2q_3}\left[|\psi^{\uparrow;qq_2q_3}_{\uparrow\uparrow\downarrow}|^2+|\psi^{\uparrow;qq_2q_3}_{\uparrow\downarrow\uparrow}|^2+|\psi^{\uparrow;qq_2q_3}_{\downarrow\uparrow\uparrow}|^2\right],\\
B^q&=\int\ud[23]\sum_{q_2q_3}\left[|\psi^{\uparrow;qq_2q_3}_{\uparrow\uparrow\downarrow}|^2+|\psi^{\uparrow;qq_2q_3}_{\uparrow\downarrow\uparrow}|^2-|\psi^{\uparrow;qq_2q_3}_{\downarrow\uparrow\uparrow}|^2\right]\nonumber\\
&=\int\ud[23]\sum_{q_2q_3}\left[\left(\psi^{\uparrow;qq_2q_3}_{\uparrow\uparrow\downarrow}\right)^*\psi^{\downarrow;qq_2q_3}_{\downarrow\uparrow\downarrow}+\left(\psi^{\uparrow;qq_2q_3}_{\uparrow\downarrow\uparrow}\right)^*\psi^{\downarrow;qq_2q_3}_{\downarrow\downarrow\uparrow}\right].
\end{align}
The TMD relations \eqref{rel1}-\eqref{rel3} then follow trivially.

\subsection{$SU(6)$ Spin-Flavor Symmetry}

Many quark models, in addition of being spherically symmetric, assume also the $SU(6)$ spin-flavor symmetry. This symmetry restricts further the components of the wave function in the canonical-spin basis in the following way
\begin{equation}
\begin{tabular}{c|ccc}
$\psi^{\uparrow;q_1q_2q_3}_{\sigma_1\sigma_2\sigma_3}$&$\,\,uud\,\,$&$\,\,udu\,\,$&$\,\,duu\,\,$\\
\hline
$\uparrow\uparrow\downarrow$&$2\phi$&$-\phi$&$-\phi$\\
$\uparrow\downarrow\uparrow$&$-\phi$&$2\phi$&$-\phi$\\
$\downarrow\uparrow\uparrow$&$-\phi$&$-\phi$&$2\phi$
\end{tabular}
\end{equation}
with $\phi=\phi(\{\tilde k_i\})$ some symmetric momentum wave function. This implies that the unpolarized and polarized amplitudes $A^q$ and $B^q$ are simply proportional
\begin{equation}
A^u=2A^d=\tfrac{3}{2}\,B^u=-6\,B^d=12\int\ud[23]|\phi|^2,
\end{equation}
and so the flavor-dependent relation \eqref{rel4} follows trivially with $\mathcal D^q=B^q/A^q$.

\section{Quark Models}\label{models}
In this section we review  different quark models which have been used for the calculation of TMDs. In particular, we summarize the main ingredients of the models and discuss whether they satisfy the conditions of Sec.~\ref{section-3}. In order to facilitate the discussion, we sort the quark models in classes defined as follows: 
\begin{itemize}
\item The light-cone constituent quark model (LCCQM) of Ref.~\cite{Pasquini:2008ax} and the light-cone quark-diquark model (LCQDM) of Refs.~\cite{She:2009jq,Zhu:2011zza} constitute the class of light-cone models;
\item The covariant parton model of Ref.~\cite{Efremov:2009ze} constitutes its own class;
\item The bag model of Refs.~\cite{Avakian:2010br,Avakian:2008dz} and the light-cone version of the chiral quark-soliton model ($\chi$QSM) of Refs.~\cite{Lorce:2011dv,Lorce:2006nq,Lorce:2007as} constitute the class of mean-field models; 
\item The quark-diquark models of Refs.~\cite{Jakob:1997wg,Goldstein:2002vv,Gamberg:2003ey,Gamberg:2007wm,Bacchetta:2008af} 
constitute the class of spectator models.
\end{itemize}
We do not discuss the quark-target model of Ref.~\cite{Meissner:2007rx} 
as it deals with gluons and therefore does already not satisfy the first condition of Sec.~\ref{section-3}.

\subsection{Light-Cone Models}

The class of light-cone models is characterized by the fact that the target state is expanded in the basis of free parton (Fock) states. One usually truncates the expansion and considers only the state with the lowest number of partons. In the LCCQM, this lowest state consists of three valence quarks, while in the LCQDM it consists of a valence quark and a spectator diquark.

It is well known that light-cone helicity and canonical spin of free partons are simply related by the so-called Melosh rotation \cite{Melosh:1974cu}. Its $j=1/2$ and $j=1$ representations \cite{Ahluwalia:1993xa} are given by (see App.~\ref{appendix:a} for the definition of the spinors and polarization four-vectors) 
\begin{align}
D^{(1/2)*}_{\sigma\lambda}(\tilde k)&=\frac{\overline u_{LC}(k,\lambda)u(k,\sigma)}{2m}\nonumber\\
&=\frac{1}{\sqrt{N}}\begin{pmatrix}\sqrt{2}\,k^++m&-k_R\\k_L&\sqrt{2}\,k^++m\end{pmatrix},\label{meloshspinor}\\
D^{(1)*}_{\sigma\lambda}(\tilde k)&=-\varepsilon^*_{LC}(k,\lambda)\cdot\varepsilon(k,\sigma)\nonumber\\
&=\frac{1}{N}\begin{pmatrix}\left(\sqrt{2}\,k^++m\right)^2&-\sqrt{2}\left(\sqrt{2}\,k^++m\right)k_R&k_R^2\\\sqrt{2}\left(\sqrt{2}\,k^++m\right)k_L&\left(\sqrt{2}\,k^++m\right)^2-\uk_\perp^2&-\sqrt{2}\left(\sqrt{2}\,k^++m\right)k_R\\k_L^2&\sqrt{2}\left(\sqrt{2}\,k^++m\right)k_L&\left(\sqrt{2}\,k^++m\right)^2\end{pmatrix},\label{meloshvector}
\end{align}
where $m$ is the parton mass and $N=(\sqrt{2}\,k^++m)^2+\uk_\perp^2$. The LCWF in the canonical-spin basis being identified in these models with the instant-form wave function, it follows that $\sqrt{2}\,k^+=x\mathcal M_0$ with $\mathcal M_0=\sum_i\omega_i$ the mass of the free parton state and $\omega_i$ the free energy of parton $i$. Comparing now Eqs.~\eqref{meloshspinor} and \eqref{meloshvector} with Eqs.~\eqref{genMelosh} and \eqref{genMelosh2}, we obtain
\begin{equation}
\cos\tfrac{\theta}{2}=\frac{m+x\mathcal M_0}{\sqrt{N}}\qquad\text{and}\qquad\sin\tfrac{\theta}{2}=\frac{k_\perp}{\sqrt{N}}.
\end{equation}

Finally, both the LCCQM and LCQDM consider wave functions with spherical symmetry and $SU(6)$ spin-flavor symmetry. In other words, all the conditions of Sec.~\ref{section-3} are satisfied in these models, and so are the TMD relations \eqref{rel1}-\eqref{rel4}.

\subsection{Covariant Parton Model}

The standard quark-parton model (QPM) refers to the infinite momentum frame (IMF), where the parton mass can be neglected. The covariant parton model is an alternative to the QPM that is not confined to a preferred reference frame. Following the standard assumptions of the QPM, the covariant parton model describes the target system as a gas of quasi-free partons, \emph{i.e.} the partons bound inside the target behave at the interaction with the external probe (at sufficiently high $Q^2$) as free particles having four-momenta on the mass shell. However, since the covariant parton model does not refer specifically to the IMF, the parton mass\footnote{Note that the parton mass appearing in the model has to be regarded as an effective mass, in the sense that it corresponds to the mass of the free parton behaving at the interaction like the actual bound parton.} $m$ is not neglected. One also assumes explicitly that the parton phase space is spherical.

The covariant parton model does not refer explicitly to quark canonical spin or light-cone helicity. Instead, it deals with the covariant quark polarization vector. Identifying in the Bjorken limit the Lorentz structures of the hadronic tensor with those of the TMD correlator, the authors of Ref.~\cite{Efremov:2009ze} found that the TMDs are given in the covariant parton model by
\begin{subequations}
\begin{align}
f^q_1&=\tfrac{1}{2}\left[\left(m+xM\right)^2+\uk^2_\perp\right]\int\{\ud\tilde k^1\},\\
g^q_{1L}&=\tfrac{1}{2}\left[\left(m+xM\right)^2-\uk^2_\perp\right]\int\{\ud k^1\},\\
\tfrac{k_\perp}{M}\,f^{\perp q}_{1T}&=0,\\
\tfrac{k_\perp}{M}\,g^q_{1T}&=\left(m+xM\right)k_\perp\int\{\ud k^1\},\\
\tfrac{k_\perp}{M}\,h^{\perp q}_1&=0,\\
\tfrac{k_\perp}{M}\,h^{\perp q}_{1L}&=-\left(m+xM\right)k_\perp\int\{\ud k^1\},\\
\tfrac{\uk_\perp^2}{2M^2}\,h^{\perp q}_{1T}&=-\tfrac{1}{2}\,\uk_\perp^2\int\{\ud k^1\},
\end{align}
\end{subequations}
where $M$ is the target mass, $\{\ud\tilde k^1\}$ and $\{\ud k^1\}$ are the measures associated to the distributions of unpolarized and polarized quarks, respectively. Comparing with Eq.~\eqref{sphericalTMDs}, we find that
\begin{equation}
\cos\tfrac{\theta}{2}=\frac{m+xM}{\sqrt{\left(m+xM\right)^2+\uk_\perp^2}}\qquad\text{and}\qquad\sin\tfrac{\theta}{2}=\frac{k_\perp}{\sqrt{\left(m+xM\right)^2+\uk_\perp^2}},
\end{equation}
which is nothing else than the Melosh rotation. This is consistent with the fact that the active quark is quasi-free in this model. The difference with light-cone models is that the physical mass of the target $M$ is used in the Melosh rotation instead of the free invariant mass\footnote{In the covariant parton model, only the active parton is considered on-shell. In light-cone models, all the partons are on-shell so that the Fock state itself is off-shell $\mathcal M_0\neq M$.} $\mathcal M_0$. The conditions 1-3 of Sec.~\ref{section-3} are therefore satisfied in the covariant parton model, and so are the TMD relations \eqref{rel1}-\eqref{rel3}. Since this model does not use the language of wave functions, the implementation of $SU(6)$ spin-flavor symmetry is more delicate and one has to assume that the unpolarized and polarized distributions become simply proportional in order to recover the TMD relation \eqref{rel4}.

\subsection{Mean-Field Models}

In mean-field models, the target is considered as made of quarks bound by a classical mean field representing the non-perturbative (long-range) contribution of the gluon field. Accordingly, the positive-frequency part of the quark field appearing in the definition of the correlator \eqref{correlator} is expanded in the basis of the bound-state solutions $e^{-iE_nt}\,\varphi_n(k,\sigma)$ instead of the free Dirac light-cone spinors $e^{-ip\cdot x}\,u_{LC}(k,\lambda)$. Moreover, one truncates the expansion to the lowest mode $\varphi\equiv\varphi_1$ with energy $E_\text{lev}\equiv E_1$.

In these models, the bound-state solution $\varphi(k,\sigma)$ is called the quark wave function. This object is clearly different from the LCWF introduced in Sec.~\ref{section-5}. 
In particular, the former is a spinor while the latter is an ordinary scalar function. It is however possible to relate them.
Since we consider twist-$2$ Dirac operators $\Gamma$, only the good components of the spinors are involved in the quark bilinear $\overline\varphi(k,\sigma')\Gamma\varphi(k,\sigma)$. Using $u^\dag_G(\lambda)=P_+u_{LC}(k,\lambda)/\sqrt{2^{1/2}k^+}$ (see App.~\ref{appendix:a}), we find that
\begin{align}
\overline\varphi(k,\sigma')\Gamma\varphi(k,\sigma)&=\varphi^\dag(k,\sigma')P_+\gamma^0\Gamma P_+\varphi(k,\sigma)\nonumber\\
&=\sum_{\lambda\lambda'}F^*_{\lambda'\sigma'}(k')\,F_{\lambda\sigma}(k)\,\frac{\overline u_{LC}(k,\lambda')\Gamma u_{LC}(k,\lambda)}{\sqrt{2}k^+},
\end{align}
where we have defined $F_{\lambda\sigma}(k)=u^\dag_G(\lambda)\varphi(k,\sigma)$. In agreement with \cite{Petrov:2002jr,Diakonov:2005ib} where one boosts explicitly the system in the mean-field approximation to the IMF, $F_{\lambda\sigma}(k)$ can be interpreted as the quark LCWF with $k_z=xM-E_\text{lev}$. The mass $M$ is identified with the nucleon mass $M_N$ in the bag model and with the soliton mass $\uM_N$ in the $\chi$QSM.

The mean field is assumed spherically symmetric in the target rest frame. It follows that the lowest quark-state solution in momentum space takes the form 
\begin{equation}\label{sphericalspinor}
\varphi(k,\sigma)=\begin{pmatrix}f(|\uk|)\\\tfrac{\uk\cdot\usigma}{|\uk|}\,g(|\uk|)\end{pmatrix}\chi_\sigma,
\end{equation}
with $\chi_\sigma$ the Pauli spinor.
The functions $f$ and $g$ in Eq.~\eqref{sphericalspinor} represent the $s$ ($\ell=0$) and $p$ ($\ell=1$) waves of the bound-state solution. On the other hand, the general 3Q LCWF for a spin-$1/2$ target involves usually $s$-, $p$- and $d$-waves. There is no contradiction between these two statements since $f$ and $g$ describe a single quark in the target and therefore do not represent partial waves of \emph{total} angular momentum. Note also that in the language of 3Q LCWF, the $s$-, $p$- and $d$-waves refer to components with $\ell_z=0$, $\pm 1$ and $\pm 2$, respectively. This is an abuse of language as partial waves should refer to $\ell$ and not $\ell_z$.

The quark LCWF corresponding to \eqref{sphericalspinor} is then given by\footnote{Replacing $\varphi(k,\sigma)$ by the free Dirac spinor $u(k,\sigma)$, one recovers the Melosh rotation given by Eq.~\eqref{meloshspinor} $u^\dag_G(\lambda)u(k,\sigma)=\sqrt{2^{1/2}k^+}\,D^{(1/2)*}_{\sigma\lambda}(\tilde k)$.}
\begin{equation}
F_{\lambda\sigma}(k)=\frac{1}{\sqrt{2}}\begin{pmatrix}f(|\uk|)+\tfrac{k_z}{|\uk|}\,g(|\uk|)&-\tfrac{k_R}{|\uk|}\,g(|\uk|)\\\tfrac{k_L}{|\uk|}\,g(|\uk|)&f(|\uk|)+\tfrac{k_z}{|\uk|}\,g(|\uk|)\end{pmatrix}_{\sigma\lambda}.
\end{equation}
It describes in particular how canonical spin $\sigma$ and light-cone helicity $\lambda$ are related\footnote{Defining the quark rest-frame wave function as in Refs.~\cite{Petrov:2002jr,Diakonov:2005ib} $F^\text{rest}_{\sigma'\sigma}(k)=u^\dag(k,\sigma')\varphi(k,\sigma)$, one finds that $F^\text{rest}_{\sigma'\sigma}(k)=\left[\sqrt{E+m}\,f(|\uk|)+\sqrt{E-m}\,g(|\uk|)\right]\delta_{\sigma'\sigma}$ which is consistent with $\sigma$ being identified with canonical spin.}. Comparing with Eq.~\eqref{genMelosh}, we find that
\begin{equation}
\cos\tfrac{\theta}{2}=\frac{f(|\uk|)+\tfrac{k_z}{|\uk|}\,g(|\uk|)}{\sqrt{N}}\qquad\text{and}\qquad\sin\tfrac{\theta}{2}=\frac{\tfrac{k_\perp}{|\uk|}\,g(|\uk|)}{\sqrt{N}},
\end{equation}
with $N=f^2(|\uk|)+2\,\tfrac{k_z}{|\uk|}\,f(|\uk|)g(|\uk|)+g^2(|\uk|)$. The 3Q LCWF written as $\prod_{i=1}^{3}F_{\lambda_i\sigma_i}(k_i)$ times the standard $SU(6)$ spin-flavor wave function with target polarization $\Lambda=\sum_i\sigma_i$, is then consistent with spherical symmetry in the canonical-spin basis. All the conditions of Section~\ref{section-3} being satisfied in mean-field models, the TMD relations \eqref{rel1}-\eqref{rel4} follow automatically.

\subsection{Spectator Models}

The basic idea of spectator models is to evaluate the quark-quark correlator $\Phi$ of Eq.~\eqref{correlator} by inserting a complete set of intermediate states and then truncating this set at tree level to a single on-shell spectator diquark state, \emph{i.e.} a state with the quantum numbers of two quarks. The diquark can be either an isospin singlet with spin $0$ (scalar diquark) or an isospin triplet with spin $1$ (axial-vector diquark). The target is then seen as made of an off-shell quark and an on-shell diquark. Spectator models differ by their specific choice of target-quark-diquark vertices, polarization four-vectors associated with the axial-vector diquark, and form factors which take into account in an effective way the composite nature of the target and the spectator diquark. 

As advocated in Ref.~\cite{Brodsky:2000ii}, the parton distributions can conveniently be computed using the language of LCWFs. The scalar quark-diquark LCWF is defined as
\begin{equation}
\psi^\Lambda_\lambda(\tilde k)\propto\overline u_{LC}(k,\lambda)\mathcal Y_s u_{LC}(P,\Lambda)
\end{equation}
with target momentum $P=\left[P^+,\tfrac{M^2}{2P^+},\uzero_\perp\right]$. We do not need to specify all the factors in the definition as we are only interested in the structure of the wave function in the light-cone helicity basis. The scalar vertex is  of the Yukawa type $\mathcal Y_s=g_s(k^2)\,\mathds 1$ with $g_s(k^2)$ some form factor. Writing down explicitly the components, one finds
\begin{equation}
\psi^\Lambda_\lambda(\tilde k)\propto\frac{g_s(k^2)}{\sqrt{x}}\begin{pmatrix}m+xM&-k_R\\k_L&m+xM\end{pmatrix}_{\Lambda\lambda}.
\label{melosh-diquark}
\end{equation}
Note the striking resemblance with the Melosh rotation matrix of Eq.~\eqref{meloshspinor}. One can similarly define a rest-frame scalar quark-diquark wave function as
\begin{equation}
\label{wf-restframe}
\Psi^\Lambda_\sigma\propto\overline u(k,\sigma)\mathcal Y_s u(P_\text{rest},\Lambda)=g_s(k^2)\sqrt{2M(E+m)}\,\delta_{\Lambda\sigma}
\end{equation}
with target momentum $P_\text{rest}=\left(M,\uzero\right)$. 
This wave function is obviously spherically symmetric
\footnote{ The rest-frame wave function in Eq.~(\ref{wf-restframe})
is expressed in terms of canonical spin and therefore has the same spin 
structure as the LCWF expressed in the canonical-spin basis. It follows that
the constraints due to spherical symmetry discussed in 
apppendix~\ref{app-c1}
apply also here.
Furthermore, the momentum dependent 
part of the wave function in the rest frame does not depend on a specific direction.
}.
Furthermore, Eq.~(\ref{melosh-diquark})
suggests that the quark light-cone helicity and canonical spin are simply related by a Melosh rotation, as if the quark was free~\cite{Ellis:2008in}. In other words, the conditions 1-3 of Section~\ref{section-3} are satisfied in scalar diquark models with Yukawa-like vertex, and so are the TMD relations \eqref{rel1}-\eqref{rel3}.

The axial-vector quark-diquark LCWF is defined as
\begin{equation}
\psi^\Lambda_{\lambda\lambda_D}(\tilde k)\propto\overline u_{LC}(k,\lambda)\varepsilon^*_{LC\mu}(K,\lambda_D)\mathcal Y^\mu_a u_{LC}(P,\Lambda).
\end{equation}
The spectator model of Jakob \emph{et al.} \cite{Jakob:1997wg} assumes the following structure for the axial-vector vertex $\mathcal Y^\mu_a=\frac{g_a(k^2)}{\sqrt{3}}\,\gamma_5\left(\gamma^\mu+\tfrac{P^\mu}{M}\right)$ and the following momentum argument for the polarization four-vector $K=P$. The motivation for such a choice is to ensure
 that, in the target rest frame, the diquark spin-$1$ states are purely spatial. Indeed, the rest-frame axial-vector quark-diquark wave function reads in this model
\begin{equation}
\Psi^\Lambda_{\sigma\sigma_D}\propto\overline u(k,\sigma)\varepsilon^*_\mu(K,\sigma_D)\mathcal Y^\mu_a u(P_\text{rest},\Lambda)=\frac{g_a(k^2)}{\sqrt{3}}\,\sqrt{2M(E+m)}\left(\boldsymbol\epsilon_{\sigma_D}\cdot\usigma\right)_{\sigma\Lambda}.
\end{equation}
It satisfies the constraints \eqref{sphericalAV} and is therefore spherically symmetric. Writing down explicitly the components of the corresponding LCWF, one finds
\begin{equation}\label{JakobAV}
\begin{gathered}
\psi^+_{+0}\propto\frac{g_a(k^2)}{\sqrt{3x}}\left(m+xM\right),\qquad\psi^+_{-0}\propto-\frac{g_a(k^2)}{\sqrt{3x}}\,k_R,\\
\psi^+_{-+}=-\sqrt{2}\,\psi^+_{+0},\qquad\psi^-_{--}=\sqrt{2}\,\psi^+_{-0},\qquad\psi^+_{+-}=\psi^+_{--}=0,
\end{gathered}
\end{equation}
the other components being given by $\psi^{-\Lambda}_{-\lambda-\lambda_D}=(-1)^{\Lambda+\lambda+\lambda_D}\left(\psi^{\Lambda}_{\lambda\lambda_D}\right)^*$, with $\Lambda,\lambda=\pm\tfrac{1}{2}$ and $\lambda_D=+1, 0, -1$. 
Again, one recognizes the characteristic factors of the Melosh rotation~\cite{Ellis:2008in}. 
Comparing the structure of the components of the LCWF in Eq.~\eqref{JakobAV} 
with the structure of the components of the LCWF given in Table \ref{AVLCWF} after applying the constraints of spherical symmetry in the canonical-spin basis \eqref{sphericalAV}, one concludes that only the quark polarization is rotated. This is in agreement with the fact that the momentum argument of the polarization four-vector $\varepsilon_\mu$ does not have any transverse momentum, and so there is no rotation of the diquark polarization. All the  conditions 1-3 of Section~\ref{section-3} being satisfied in the axial-vector diquark model of Ref.~\cite{Jakob:1997wg}, the TMD relations \eqref{rel1}-\eqref{rel3} follow automatically. The flavor-dependent relation \eqref{rel4} can be obtained by further imposing $SU(6)$ spin-flavor symmetry to the wave function\footnote{The scalar and axial-vector diquarks represent in principle more than just two quarks. For this reason, they have \emph{a priori} different masses, cutoffs, form factors, \ldots When we impose $SU(6)$ symmetry, we implicitly consider that the quark-diquark picture originates from a 3Q picture. The scalar and axial-vector diquarks then just differ by their spin and flavor structures which are uniquely determined by the $SU(6)$ symmetry.}.

On the contrary, some versions of the spectator model presented by  Bacchetta \emph{et al.} in Ref.~\cite{Bacchetta:2008af} do not support any TMD relation. We therefore expect that at least one of the conditions 1-3
of Sec.~\ref{section-3} is not satisfied. These versions are based on the axial-vector vertex $\mathcal Y^\mu_a=\frac{g_a(k^2)}{\sqrt{2}}\,\gamma^\mu\gamma_5$ and involve the diquark momentum $K=P-k$ in the polarization four-vector.
With these choices, it is found that the condition 3 of Sec.~\ref{section-3} is not fulfilled since the corresponding rest-frame wave function  does not satisfy the requirements of spherical symmetry
\begin{equation}
\Psi^\Lambda_{\sigma\sigma_D}\not\propto\left(\boldsymbol\epsilon_{\sigma_D}\cdot\usigma\right)_{\sigma\Lambda},
\end{equation}
in accordance with the discussions of Refs.~\cite{Gross:2006fg,Ramalho:2008ra,Gross:2008zza} and the comment in Ref.~\cite{Bacchetta:2008af} that in this approach the partons do not necessarily occupy the lowest-energy available orbital (with quantum numbers $J^P=\tfrac{1}{2}^+$ and $L_z=0$.)

\section{Conclusions}

In this work we presented a study of 
the transverse-momentum dependent parton distributions in the framework of quark models.
We focused the discussion on model relations which appeared in a large panel
of quark models, elucidating their physical origin and implications.
In particular, there are in total four independent relations among the leading-twist TMDs: 
three of them are flavor independent and connect polarized TMDs,  while a fourth flavor-dependent relation involves both polarized and unpolarized TMDs.
\\
We have shown that these model relations have essentially a geometrical origin,
and can be traced back to properties of rotational invariance of the system.
 In particular, 
we identified the  conditions which are sufficient for the existence of the flavor-independent relations.
They are:
\begin{enumerate}
\item 
the probed quark behaves as if it does not interact directly with the other partons ({\em i.e.} one works within the standard impulse approximation) and
there are no explicit gluons;
\item the quark light-cone and canonical polarizations are related by a rotation with axis orthogonal to both the light-cone and quark transverse-momentum directions;
\item the target has spherical symmetry in the canonical-spin basis.
\end{enumerate}
For the flavor-dependent relation, one  needs a further condition for the spin-flavor dependent part
of the nucleon wave function. Specifically, it is required 
\begin{enumerate}[resume]
\item  $SU(6)$ spin-flavor symmetry of the wave function.
\end{enumerate}
On the basis of the above assumptions, we were able to derive the model relations among TMDs within two different approaches.

The first approach is based on the representation of the quark correlator entering the definition of TMDs 
in terms of the polarization amplitudes of the quarks and nucleon. 
Such amplitudes are usually expressed in the basis of light-cone helicity.
However, in order to discuss in a simple way the rotational  properties of the system,
we introduced the representation in the basis of canonical spin. 
In this framework, we showed that 
the conditions 1-3 are sufficient for the existence 
of all three flavor-independent relations.
We also showed that a subset of these three relations can be derived relaxing the assumption
 of spherical symmetry and using the less restrictive condition of axial symmetry under a rotation around a specific direction.

The second approach is based on the representation of TMDs in terms of quark wave functions.
In particular, we expressed the TMDs as overlap of light-cone wave functions, 
and we derived the relation with the corresponding representation in terms of overlap of 
wave functions in the canonical-spin basis.
After discussing the consequence of spherical symmetry on the spin structure of the wave function, we 
were able to obtain  an alternative derivation  of the relations among polarized TMDs.
Finally, for the remaining relation among polarized and unpolarized TMDs, 
we used the 
$SU(6)$ symmetry for the spin-isospin dependence of the nucleon wave function.

The previous formal discussion 
has been made more concrete with examples from quark models which have been used in literature.
Besides the specific assumptions for modeling the quark dynamics, these models can be sorted in different classes corresponding to light-cone models, the covariant parton models, mean-field models and spectator models.
We have  shown how and to which extent the conditions 1-4 are realized in these different models.
In particular we verified that all the models satisfying the TMD relations also satisfy the above conditions, while models where the TMD relations do not hold fail with at least one of the above conditions.

Finally, we remark that these relations are not expected to hold identically in QCD where TMDs are all independent.
However,
 they provide simplified and intuitive notions for the interpretation of the spin and
orbital angular momentum structure of the nucleon. 
As such, they can be useful for phenomenological studies to build up simplified parametrizations of TMDs to be fitted to  data. Furthermore, the comparison with the experimental data will tell us the degree of accuracy of such relations, giving insights for further studies 
towards more refined
 quark models. 

\section*{Acknowledgments}

C. L. is thankful to INFN and the Department of Nuclear and Theoretical Physics of the University of Pavia for the hospitality. We also
acknowledge  very kind and instructive discussions with A. Bacchetta and P. Schweitzer. This work was supported in part by the Research Infrastructure Integrating Activity ``Study of Strongly Interacting Matter'' (acronym HadronPhysics2, Grant Agreement n. 227431) under the Seventh Framework Programme of the European Community, by the Italian MIUR through the PRIN 2008EKLACK ``Structure of the nucleon: transverse momentum, transverse spin and orbital angular momentum''.

\appendix

\section{Spinors and Polarization Four-Vectors}
\label{appendix:a}
We collect in this Appendix the different types of free spinors and polarization vectors. The free canonical Dirac spinor $u(k,\sigma)$ and polarization four-vector $\varepsilon^\mu(k,\sigma)$ are given by
\begin{align}
u(k,\sigma)&=\begin{pmatrix}\sqrt{E+m}\,\mathds{1}\\\sqrt{E-m}\,\frac{\uk\cdot\usigma}{|\uk|}\end{pmatrix}\chi_\sigma,\\
\varepsilon^\mu(k,\sigma)&=\left(\frac{\boldsymbol\epsilon_\sigma\cdot\uk}{m},\boldsymbol\epsilon_\sigma+\frac{\uk\left(\boldsymbol\epsilon_\sigma\cdot\uk\right)}{m(E+m)}\right),
\end{align}
where $\chi_\uparrow=\left(\begin{smallmatrix}1\\0\end{smallmatrix}\right)$, $\chi_\downarrow=\left(\begin{smallmatrix}0\\1\end{smallmatrix}\right)$, 
and the polarization  three-vectors are  $\boldsymbol\epsilon_{\Uparrow,\Downarrow}=\frac{1}{\sqrt{2}}\left(\mp 1,-i,0\right)$ for $s_z=\pm 1$, and $\boldsymbol\epsilon_\odot=\left(0,0,1\right)$ for $s_z=0$.
The free light-cone Dirac spinor $u_{LC}(k,\lambda)$ and polarization four-vector $\varepsilon^\mu_{LC}(k,\lambda)$ are given by
\begin{align}
u_{LC}(k,+)&=\frac{1}{\sqrt{2^{3/2}k^+}}\begin{pmatrix}\sqrt{2}\,k^++m\\k_R\\\sqrt{2}\,k^+-m\\k_R\end{pmatrix}, &u_{LC}(k,-)&=\frac{1}{\sqrt{2^{3/2}k^+}}\begin{pmatrix}-k_L\\\sqrt{2}\,k^++m\\k_L\\-\sqrt{2}\,k^++m\end{pmatrix},\\
\varepsilon^\mu_{LC}(k,\pm)&=\left[0,\frac{\boldsymbol\epsilon_{\perp\pm}\cdot\uk_\perp}{k^+},\boldsymbol\epsilon_{\perp\pm}\right], &\varepsilon^\mu_{LC}(k,0)&=\frac{1}{m}\left[k^+,\frac{\uk_\perp^2-m^2}{2k^+},\uk_\perp\right],
\end{align}
with $\boldsymbol\epsilon_{\perp\pm}=\tfrac{1}{\sqrt{2}}\left(\mp 1,-i\right)$. Both types of spinors and polarization four-vectors coincide in the rest frame $k_\text{rest}=(m,\uzero)$
\begin{align}
u(k_\text{rest},\sigma)&=u_{LC}(k_\text{rest},\sigma)=\sqrt{2m}\begin{pmatrix}\chi_\sigma\\0\end{pmatrix},\\
\varepsilon^\mu(k_\text{rest},\sigma)&=\varepsilon^\mu_{LC}(k_\text{rest},\sigma)=\left(0,\boldsymbol\epsilon_\sigma\right).
\end{align}

The ``good'' light-cone spinors are the simultaneous eigenstates of the operator $\gamma_5$ and the projector $P_+=\tfrac{1}{2}\,\gamma^-\gamma^+$ 
\begin{equation}
P_+u_G(\lambda)=u_G(\lambda),\qquad \gamma_5u_G(\lambda)=\lambda\,u_G(\lambda),\qquad u_G(\lambda)\equiv\frac{1}{\sqrt{2}}\begin{pmatrix}\mathds{1}\\\sigma_3\end{pmatrix}\chi_\lambda,
\end{equation}
and one can write
\begin{equation}
P_+=\sum_\lambda u_G(\lambda)u^\dag_G(\lambda).
\end{equation}

\section{Components of the 3Q LCWF in the Light-Cone and Canonical Polarization Bases}\label{table}

Based on Eq.~\eqref{connection}, Table~\ref{3QLCWF} shows explicitly how the components of the 3Q LCWF in light-cone polarization basis are decomposed in the canonical-spin basis. 
\begin{table}[h!]
\begin{center}
\caption{\footnotesize{Decomposition in the canonical-spin basis $\psi^\uparrow_{\sigma_1\sigma_2\sigma_3}$ of the components of the 3Q LCWF in the light-cone helicity basis $\psi^+_{\lambda_1\lambda_2\lambda_3}$. The components are grouped according to the values of total orbital angular momentum $\ell_z$.\vspace{.4cm}}}\label{3QLCWF}
\begin{tabular}{lc||c|ccc|ccc|c}
&&$\,\,\ell_z=-1\,\,$&\multicolumn{3}{c|}{$\ell_z=0$}&\multicolumn{3}{c|}{$\ell_z=+1$}&$\,\,\ell_z=+2\,\,$\\
&&$\psi^\uparrow_{\uparrow\uparrow\uparrow}$&$\psi^\uparrow_{\uparrow\uparrow\downarrow}$&$\psi^\uparrow_{\uparrow\downarrow\uparrow}$&$\psi^\uparrow_{\downarrow\uparrow\uparrow}$&$\psi^\uparrow_{\downarrow\downarrow\uparrow}$&$\psi^\uparrow_{\downarrow\uparrow\downarrow}$&$\psi^\uparrow_{\uparrow\downarrow\downarrow}$&$\psi^\uparrow_{\downarrow\downarrow\downarrow}$\\
\hline\hline
$\ell_z=-1\quad$&$\psi^+_{+++}$&$z_1z_2z_3$&$z_1z_2l_3$&$z_1l_2z_3$&$l_1z_2z_3$&$l_1l_2z_3$&$l_1z_2l_3$&$z_1l_2l_3$&$l_1l_2l_3$\\\hline
&$\psi^+_{++-}$&$-z_1z_2r_3$&$z_1z_2z_3$&$-z_1l_2r_3$&$-l_1z_2r_3$&$-l_1l_2r_3$&$l_1z_2z_3$&$z_1l_2z_3$&$l_1l_2z_3$\\
$\ell_z=0$&$\psi^+_{+-+}$&$-z_1r_2z_3$&$-z_1r_2l_3$&$z_1z_2z_3$&$-l_1r_2z_3$&$l_1z_2z_3$&$-l_1r_2l_3$&$z_1z_2l_3$&$l_1z_2l_3$\\
&$\psi^+_{-++}$&$-r_1z_2z_3$&$-r_1z_2l_3$&$-r_1l_2z_3$&$z_1z_2z_3$&$z_1l_2z_3$&$z_1z_2l_3$&$-r_1l_2l_3$&$z_1l_2l_3$\\\hline
&$\psi^+_{--+}$&$r_1r_2z_3$&$r_1r_2l_3$&$-r_1z_2z_3$&$-z_1r_2z_3$&$z_1z_2z_3$&$-z_1r_2l_3$&$-r_1z_2l_3$&$z_1z_2l_3$\\
$\ell_z=+1$&$\psi^+_{-+-}$&$r_1z_2r_3$&$-r_1z_2z_3$&$r_1l_2r_3$&$-z_1z_2r_3$&$-z_1l_2r_3$&$z_1z_2z_3$&$-r_1l_2z_3$&$z_1l_2z_3$\\
&$\psi^+_{+--}$&$z_1r_2r_3$&$-z_1r_2z_3$&$-z_1z_2r_3$&$l_1r_2r_3$&$-l_1z_2r_3$&$-l_1r_2z_3$&$z_1z_2z_3$&$l_1z_2z_3$\\\hline
$\ell_z=+2$&$\psi^+_{---}$&$-r_1r_2r_3$&$r_1r_2z_3$&$r_1z_2r_3$&$z_1r_2r_3$&$-z_1z_2r_3$&$-z_1r_2z_3$&$-r_1z_2z_3$&$z_1z_2z_3$\\
\end{tabular}
\end{center}
\end{table}
We used for convenience the notations $z_i=\cos\tfrac{\theta_i}{2}$, $l_i=\hat k_{iL}\,\sin\tfrac{\theta_i}{2}$ and $r_i=\hat k_{iR}\,\sin\tfrac{\theta_i}{2}$ for the components of the rotation matrix $D^{(1/2)*}_{\sigma_i\lambda_i}$ of Eq.~\eqref{genMelosh}. For example, from the first row of Table~\ref{3QLCWF}, we have
\begin{multline}
\qquad\psi^+_{+++}=z_1z_2z_3\,\psi^\uparrow_{\uparrow\uparrow\uparrow}+z_1z_2l_3\,\psi^\uparrow_{\uparrow\uparrow\downarrow}+z_1l_2z_3\,\psi^\uparrow_{\uparrow\downarrow\uparrow}+l_1z_2z_3\,\psi^\uparrow_{\downarrow\uparrow\uparrow}\\
+l_1l_2z_3\,\psi^\uparrow_{\downarrow\downarrow\uparrow}+l_1z_2l_3\,\psi^\uparrow_{\downarrow\uparrow\downarrow}+z_1l_2l_3\,\psi^\uparrow_{\uparrow\downarrow\downarrow}+l_1l_2l_3\,\psi^\uparrow_{\downarrow\downarrow\downarrow}.\qquad
\end{multline}
It is interesting to note that any single component of the 3Q LCWF in the canonical-spin basis contributes to all components in the light-cone helicity basis, and \emph{vice-versa}. So even if one considers that the wave function has only components with $\ell_z=0$ in the canonical-spin basis, the components of the wave function in the light-cone helicity basis present all the values $\ell_z=-1,0,+1,+2$, the orbital angular momentum being generated by the rotation matrices $D^{(1/2)*}_{\sigma_i\lambda_i}$.

\section{Connection to a Quark-Diquark Picture}

We show in this Appendix how the 3Q picture can be connected to a quark-diquark picture. In the latter, one considers the whole spectator system as an object with the quantum numbers of two quarks, namely a diquark. One may also  assume that this diquark does not contain any internal orbital angular momentum. From a 3Q picture, this amounts to set $\tilde k_2=\tilde k_3=\tilde k_D/2$ and $m_D=2m$ with $\tilde k_D$ and $m_D$ the light-cone momentum and mass of the diquark, and $m$ the mass of a valence quark.

\subsection{Scalar Diquark}
\label{app-c1}
The scalar diquark is obtained by coupling the two spectator quarks so to form a system with total angular momentum $j=0$. The LCWF of the scalar quark-diquark system can be written in terms of the 3Q LCWF as follows
\begin{equation}
\psi^\Lambda_\lambda(\tilde k,\tilde k_D)=\tfrac{1}{\sqrt{2}}\left[\psi^\Lambda_{\lambda+-}(\tilde k,\tfrac{\tilde k_D}{2},\tfrac{\tilde k_D}{2})-\psi^\Lambda_{\lambda-+}(\tilde k,\tfrac{\tilde k_D}{2},\tfrac{\tilde k_D}{2})\right].
\end{equation}
The total orbital angular momentum of a given component $\psi^\Lambda_\lambda$ is given by the expression $\ell_z=\Lambda-\lambda$ with $\Lambda,\lambda=\pm\tfrac{1}{2}$. 

The corresponding LCWF in the canonical-spin basis is defined through
\begin{equation}\label{connectionS}
\psi^\Lambda_\lambda=\sum_\sigma\psi^\Lambda_\sigma\,D^{(1/2)*}_{\sigma\lambda},
\end{equation}
and can consistently be written as
\begin{equation}\label{dualS}
\psi^\Lambda_\sigma(\tilde k,\tilde k_D)=\tfrac{1}{\sqrt{2}}\left[\psi^\Lambda_{\sigma\uparrow\downarrow}(\tilde k,\tfrac{\tilde k_D}{2},\tfrac{\tilde k_D}{2})-\psi^\Lambda_{\sigma\downarrow\uparrow}(\tilde k,\tfrac{\tilde k_D}{2},\tfrac{\tilde k_D}{2})\right].
\end{equation}
The explicit decomposition of Eq.~\eqref{connectionS} is displayed in Table~\ref{SLCWF}. 
\begin{table}[h!]
\begin{center}
\caption{\footnotesize{Decomposition in the canonical-spin basis $\psi^\uparrow_\sigma$ of the components of the scalar quark-diquark LCWF in the light-cone helicity basis $\psi^+_\lambda$. The components are grouped according to the values of total orbital angular momentum $\ell_z$.\vspace{.4cm}}}\label{SLCWF}
\begin{tabular}{lc||c|c}
&&$\,\,\ell_z=0\,\,$&$\,\,\ell_z=+1\,\,$\\
&&$\psi^\uparrow_\uparrow$&$\psi^\uparrow_\downarrow$\\
\hline\hline
$\ell_z=0$&$\psi^+_+$&$z$&$l$\\\hline
$\ell_z=+1\quad$&$\psi^+_-$&$-r$&$z$
\end{tabular}
\end{center}
\end{table}

Spherical symmetry in the canonical-spin basis reads
\begin{equation}
\sum_{\Lambda'\sigma'}\left[u(\theta,\phi)\right]_{\sigma\sigma'}\left[u(\theta,\phi)\right]^*_{\Lambda\Lambda'}\psi^{\Lambda'}_{\sigma'}=\psi^\Lambda_\sigma
\end{equation}
and in particular implies
\begin{subequations}\label{sphericalS}
\begin{align}
\psi^{-\Lambda}_{-\sigma}&=(-1)^{\Lambda-\sigma}\,\psi^\Lambda_\sigma,\\
\psi^\uparrow_\downarrow&=0,
\end{align}
\end{subequations}
in agreement with Eqs.~\eqref{rotinv1}, \eqref{rotinv2} and \eqref{dualS}.

\subsection{Axial-Vector Diquark}

The axial-vector diquark is obtained by coupling the two spectator quarks so to form a system with total angular momentum $j=1$. The LCWF of the axial-vector quark-diquark system can be written in terms of the 3Q LCWF as follows
\begin{equation}
\begin{split}
\psi^\Lambda_{\lambda+}(\tilde k,\tilde k_D)&=\psi^\Lambda_{\lambda++}(\tilde k,\tfrac{\tilde k_D}{2},\tfrac{\tilde k_D}{2}),\\
\psi^\Lambda_{\lambda 0}(\tilde k,\tilde k_D)&=\tfrac{1}{\sqrt{2}}\left[\psi^\Lambda_{\lambda+-}(\tilde k,\tfrac{\tilde k_D}{2},\tfrac{\tilde k_D}{2})+\psi^\Lambda_{\lambda-+}(\tilde k,\tfrac{\tilde k_D}{2},\tfrac{\tilde k_D}{2})\right],\\
\psi^\Lambda_{\lambda-}(\tilde k,\tilde k_D)&=\psi^\Lambda_{\lambda--}(\tilde k,\tfrac{\tilde k_D}{2},\tfrac{\tilde k_D}{2}).
\end{split}
\end{equation}
The total orbital angular momentum of a given component $\psi^\Lambda_{\lambda\lambda_D}$ is given by the expression $\ell_z=\Lambda-\lambda-\lambda_D$ with $\Lambda,\lambda=\pm\tfrac{1}{2}$ and $\lambda_D=+1,0,-1$. 

The corresponding LCWF in the canonical-spin basis is defined through
\begin{equation}\label{connectionAV}
\psi^\Lambda_{\lambda\lambda_D}=\sum_{\sigma\sigma_D}\psi^\Lambda_{\sigma\sigma_D}\,D^{(1/2)*}_{\sigma\lambda}\,D^{(1)*}_{\sigma_D\lambda_D},
\end{equation}
with the rotation for the axial-vector diquark given by
\begin{equation}
D^{(1)*}_{\sigma_D\lambda_D}(\tilde k_D)=\begin{pmatrix}\frac{1+\cos\theta_D}{2}&-\frac{\hat k_R}{\sqrt{2}}\,\sin\theta_D&\hat k_R^2\,\frac{1-\cos\theta_D}{2}\\
\frac{\hat k_L}{\sqrt{2}}\,\sin\theta_D&\cos\theta_D&-\frac{\hat k_R}{\sqrt{2}}\,\sin\theta_D\\
\hat k_L^2\,\frac{1-\cos\theta_D}{2}&\frac{\hat k_L}{\sqrt{2}}\,\sin\theta_D&\frac{1+\cos\theta_D}{2}
\end{pmatrix},
\end{equation}
or equivalently
\begin{equation}\label{genMelosh2}
D^{(1)*}_{\sigma_D\lambda_D}(\tilde k_D)=\begin{pmatrix}\cos^2\tfrac{\theta_D}{2}&-\sqrt{2}\,\hat k_R\,\sin\tfrac{\theta_D}{2}\cos\tfrac{\theta_D}{2}&\hat k_R^2\,\sin^2\tfrac{\theta_D}{2}\\
\sqrt{2}\,\hat k_L\,\sin\tfrac{\theta_D}{2}\cos\tfrac{\theta_D}{2}&\cos^2\tfrac{\theta_D}{2}-\sin^2\tfrac{\theta_D}{2}&-\sqrt{2}\,\hat k_R\,\sin\tfrac{\theta_D}{2}\cos\tfrac{\theta_D}{2}\\
\hat k_L^2\,\sin^2\tfrac{\theta_D}{2}&\sqrt{2}\,\hat k_L\,\sin\tfrac{\theta_D}{2}\cos\tfrac{\theta_D}{2}&\cos^2\tfrac{\theta_D}{2}
\end{pmatrix}.
\end{equation}
Provided that $\theta_D(\tilde k_D)=\theta(\tilde k_D/2)$, we can consistently write the axial-vector quark-diquark LCWF in the canonical-spin basis as
\begin{equation}\label{dualAV}
\begin{split}
\psi^\Lambda_{\sigma\Uparrow}(\tilde k,\tilde k_D)&=\psi^\Lambda_{\sigma\uparrow\uparrow}(\tilde k,\tfrac{\tilde k_D}{2},\tfrac{\tilde k_D}{2}),\\
\psi^\Lambda_{\sigma\odot}(\tilde k,\tilde k_D)&=\tfrac{1}{\sqrt{2}}\left[\psi^\Lambda_{\sigma\uparrow\downarrow}(\tilde k,\tfrac{\tilde k_D}{2},\tfrac{\tilde k_D}{2})+\psi^\Lambda_{\sigma\downarrow\uparrow}(\tilde k,\tfrac{\tilde k_D}{2},\tfrac{\tilde k_D}{2})\right],\\
\psi^\Lambda_{\sigma\Downarrow}(\tilde k,\tilde k_D)&=\psi^\Lambda_{\sigma\downarrow\downarrow}(\tilde k,\tfrac{\tilde k_D}{2},\tfrac{\tilde k_D}{2}).
\end{split}
\end{equation}
The explicit decomposition of Eq.~\eqref{connectionAV} is displayed in Table~\ref{AVLCWF}.
\begin{table}[h!]
\begin{center}
\caption{\footnotesize{Decomposition in the canonical-spin basis $\psi^\uparrow_{\sigma\sigma_D}$ of the components of the axial-vector quark-diquark LCWF in the light-cone helicity basis $\psi^+_{\lambda\lambda_D}$. The components are grouped according to the values of total orbital angular momentum $\ell_z$.\vspace{.4cm}}}\label{AVLCWF}
\begin{tabular}{lc||c|cc|cc|c}
&&$\,\,\ell_z=-1\,\,$&\multicolumn{2}{c|}{$\ell_z=0$}&\multicolumn{2}{c|}{$\,\,\ell_z=+1\,\,$}&$\,\,\ell_z=+2\,\,$\\
&&$\psi^\uparrow_{\uparrow\Uparrow}$&$\psi^\uparrow_{\uparrow\odot}$&$\psi^\uparrow_{\downarrow\Uparrow}$&$\psi^\uparrow_{\downarrow\odot}$&$\psi^\uparrow_{\uparrow\Downarrow}$&$\psi^\uparrow_{\downarrow\Downarrow}$\\
\hline\hline
$\ell_z=-1\quad$&$\psi^+_{++}$&$zz^2_D$&$\sqrt{2}\,zz_Dl_D$&$lz^2_D$&$\sqrt{2}\,lz_Dl_D$&$zl^2_D$&$ll^2_D$\\\hline
\multirow{2}{*}{$\ell_z=0$}&$\psi^+_{+0}$&$-\sqrt{2}\,zz_Dr_D$&$z\left(z^2_D-r_Dl_D\right)$&$-\sqrt{2}\,lz_Dr_D$&$\,l\left(z^2_D-r_Dl_D\right)$&$\sqrt{2}\,zz_Dl_D$&$\sqrt{2}\,lz_Dl_D$\\
&$\psi^+_{-+}$&$-rz^2_D$&$-\sqrt{2}\,rz_Dl_D$&$zz^2_D$&$\sqrt{2}\,zz_Dl_D$&$-rl^2_D$&$zl^2_D$\\\hline
\multirow{2}{*}{$\ell_z=+1$}&$\psi^+_{-0}$&$\sqrt{2}\,rz_Dr_D$&$-r\left(z^2_D-r_Dl_D\right)$&$-\sqrt{2}\,zz_Dr_D$&$\,z\left(z^2_D-r_Dl_D\right)$&$-\sqrt{2}\,rz_Dl_D$&$\sqrt{2}\,zz_Dl_D$\\
&$\psi^+_{+-}$&$zr^2_D$&$-\sqrt{2}\,zz_Dr_D$&$lr^2_D$&$-\sqrt{2}\,lz_Dr_D$&$zz^2_D$&$lz^2_D$\\\hline
$\ell_z=+2$&$\psi^+_{--}$&$-rr^2_D$&$\sqrt{2}\,rz_Dr_D$&$zr^2_D$&$-\sqrt{2}\,zz_Dr_D$&$-rz^2_D$&$zz^2_D$
\end{tabular}
\end{center}
\end{table}

Spherical symmetry in the canonical-spin basis reads
\begin{equation}
\sum_{\Lambda'\sigma'\sigma'_D}\left[u(\theta,\phi)\right]_{\sigma\sigma'}\left[U(\theta,\phi)\right]_{\sigma_D\sigma'_D}\left[u(\theta,\phi)\right]^*_{\Lambda\Lambda'}\psi^{\Lambda'}_{\sigma'\sigma'_D}=\psi^\Lambda_{\sigma\sigma_D},
\end{equation}
where
\begin{equation}
U(\theta,\phi)=\begin{pmatrix}\frac{1+\cos\theta}{2}\,e^{-i\phi}&-\frac{1}{\sqrt{2}}\,\sin\theta\,e^{-i\phi}&\frac{1-\cos\theta}{2}\,e^{-i\phi}\\
\frac{1}{\sqrt{2}}\,\sin\theta&\cos\theta&-\frac{1}{\sqrt{2}}\,\sin\theta\\
\frac{1-\cos\theta}{2}\,e^{i\phi}&\frac{1}{\sqrt{2}}\,\sin\theta\,e^{i\phi}&\frac{1+\cos\theta}{2}\,e^{i\phi}
\end{pmatrix},
\end{equation}
and in particular implies 
\begin{subequations}\label{sphericalAV}
\begin{align}
\psi^{-\Lambda}_{-\sigma-\sigma_D}&=(-1)^{\Lambda+\sigma+\sigma_D}\,\psi^\Lambda_{\sigma\sigma_D},\\
\psi^\uparrow_{\uparrow\Uparrow}&=\psi^\uparrow_{\downarrow\odot}=\psi^\uparrow_{\uparrow\Downarrow}=\psi^\uparrow_{\downarrow\Downarrow}=0,\\
\psi^\uparrow_{\downarrow\Uparrow}&=-\sqrt{2}\,\psi^\uparrow_{\uparrow\odot},
\end{align}
\end{subequations}
in agreement with Eqs.~\eqref{rotinv1}-\eqref{rotinv3} and \eqref{dualAV}.


\begin{thebibliography}{99}


\bibitem{Barone:2010zz}
  V.~Barone, F.~Bradamante and A.~Martin,
  %``Transverse-spin and transverse-momentum effects in high-energy processes,''
  Prog.\ Part.\ Nucl.\ Phys.\  {\bf 65}, 267 (2010).
  %%[arXiv:1011.0909 [hep-ph]].
  %%CITATION = PPNPD,65,267;%%

\bibitem{Jakob:1997wg}
  R.~Jakob, P.~J.~Mulders and J.~Rodrigues,
  %``Modelling quark distribution and fragmentation functions,''
  Nucl.\ Phys.\  A {\bf 626}, 937 (1997).
  %[arXiv:hep-ph/9704335].
  %%CITATION = NUPHA,A626,937;%%

\bibitem{Goldstein:2002vv}
  G.~R.~Goldstein and L.~Gamberg,
  %``Transversity and meson photoproduction,''
  arXiv:hep-ph/0209085.
  %%CITATION = HEP-PH/0209085;%%

\bibitem{Pobylitsa:2002fr}
  P.~V.~Pobylitsa,
  %``T-odd quark distributions: QCD versus chiral models,''
  arXiv:hep-ph/0212027.
  %%CITATION = HEP-PH/0212027;%%

\bibitem{Gamberg:2003ey}
  L.~P.~Gamberg, G.~R.~Goldstein and K.~A.~Oganessyan,
  %``Novel transversity properties in semi-inclusive deep inelastic
  %scattering,''
  Phys.\ Rev.\  D {\bf 67}, 071504 (2003).
  %%[arXiv:hep-ph/0301018].
  %%CITATION = PHRVA,D67,071504;%%

\bibitem{Lu:2004au}
  Z.~Lu and B.~Q.~Ma,
  %``Sivers function in light-cone quark model and azimuthal spin  asymmetries
  %in pion electroproduction,''
  Nucl.\ Phys.\  A {\bf 741}, 200 (2004).
  %%[arXiv:hep-ph/0406171].
  %%CITATION = NUPHA,A741,200;%%

\bibitem{Cherednikov:2006zn}
  I.~O.~Cherednikov, U.~D'Alesio, N.~I.~Kochelev and F.~Murgia,
  %``Instanton contribution to the Sivers function,''
  Phys.\ Lett.\  B {\bf 642}, 39 (2006).
  %%[arXiv:hep-ph/0606238].
  %%CITATION = PHLTA,B642,39;%%
%\cite{Lu:2006kt}


\bibitem{Lu:2006kt}
  Z.~Lu and I.~Schmidt,
  %``Connection between the Sivers function and the anomalous magnetic
  %moment,''
  Phys.\ Rev.\  D {\bf 75}, 073008 (2007).
  %%[arXiv:hep-ph/0611158].
  %%CITATION = PHRVA,D75,073008;%%

\bibitem{Gamberg:2007wm}
  L.~P.~Gamberg, G.~R.~Goldstein and M.~Schlegel,
  %``Transverse Quark Spin Effects and the Flavor Dependence of the Boer-Mulders
  %Function,''
  Phys.\ Rev.\  D {\bf 77}, 094016 (2008).
  %%[arXiv:0708.0324 [hep-ph]].
  %%CITATION = PHRVA,D77,094016;%%



\bibitem{Pasquini:2009bv}
  B.~Pasquini, S.~Boffi and P.~Schweitzer,
  %``The spin structure of the nucleon in light-cone quark models,''
  Mod.\ Phys.\ Lett.\  A {\bf 24}, 2903 (2009).
  %%[arXiv:0910.1677 [hep-ph]].
  %%CITATION = MPLAE,A24,2903;%%

\bibitem{Efremov:2009ze}
  A.~V.~Efremov, P.~Schweitzer, O.~V.~Teryaev and P.~Zavada,
  %``Transverse momentum dependent distribution functions in a covariant parton
  %model approach with quark orbital motion,''
  Phys.\ Rev.\  D {\bf 80}, 014021 (2009).
  %[arXiv:0903.3490 [hep-ph]].
  %%CITATION = PHRVA,D80,014021;%%

\bibitem{She:2009jq}
  J.~She, J.~Zhu and B.~Q.~Ma,
  %``Pretzelosity $h_{1T}^{\perp}$ and quark orbital angular momentum,''
  Phys.\ Rev.\  D {\bf 79}, 054008 (2009).
  %%[arXiv:0902.3718 [hep-ph]].
  %%CITATION = PHRVA,D79,054008;%%

\bibitem{Zhu:2011zza}
  J.~Zhu and B.~Q.~Ma,
  %``Proposal for measuring new transverse momentum dependent parton
  %distributions g(1T) and h perp (1L) through semi-inclusive deep inelastic
  %scattering,''
  Phys.\ Lett.\  B {\bf 696}, 246 (2011).
  %%CITATION = PHLTA,B696,246;%%

\bibitem{Lu:2010dt}
  Z.~Lu and I.~Schmidt,
  %``Orbital structure of quarks inside the nucleon in the light-cone diquark
  %model,''
  Phys.\ Rev.\  D {\bf 82}, 094005 (2010).
  %%[arXiv:1008.2684 [hep-ph]].
  %%CITATION = PHRVA,D82,094005;%%

\bibitem{Avakian:2009jt}
  H.~Avakian, A.~V.~Efremov, P.~Schweitzer, O.~V.~Teryaev, F.~Yuan and P.~Zavada,
  %``Insights on non-perturbative aspects of TMDs from models,''
  Mod.\ Phys.\ Lett.\  A {\bf 24}, 2995 (2009).
  %%[arXiv:0910.3181 [hep-ph]].
  %%CITATION = MPLAE,

\bibitem{Courtoy:2008dn}
  A.~Courtoy, S.~Scopetta and V.~Vento,
  %``Model calculations of the Sivers function satisfying the Burkardt Sum
  %Rule,''
  Phys.\ Rev.\  D {\bf 79}, 074001 (2009).
  %%[arXiv:0811.1191 [hep-ph]].
  %%CITATION = PHRVA,D79,074001;%%

\bibitem{Courtoy:2009pc}
  A.~Courtoy, S.~Scopetta and V.~Vento,
  %``Analyzing the Boer-Mulders function within different quark models,''
  Phys.\ Rev.\  D {\bf 80}, 074032 (2009).
  %%[arXiv:0909.1404 [hep-ph]].
  %%CITATION = PHRVA,D80,074032;%%

\bibitem{Courtoy:2008vi}
  A.~Courtoy, F.~Fratini, S.~Scopetta and V.~Vento,
  %``A quark model analysis of the Sivers function,''
  Phys.\ Rev.\  D {\bf 78}, 034002 (2008).
  %%[arXiv:0801.4347 [hep-ph]].
  %%CITATION = PHRVA,D78,034002;%%


\bibitem{Lorce:2007fa}
  C.~Lorc\'e,
  %``Tensor charges of light baryons in the Infinite Momentum Frame,''
  Phys.\ Rev.\  D {\bf 79}, 074027 (2009). 
  %%[arXiv:0708.4168 [hep-ph]].
  %%CITATION = PHRVA,D79,074027;%%

\bibitem{Pasquini:2008ax}
  B.~Pasquini, S.~Cazzaniga and S.~Boffi,
  %``Transverse momentum dependent parton distributions in a light-cone quark
  %model,''
  Phys.\ Rev.\  D {\bf 78}, 034025 (2008).
  %%[arXiv:0806.2298 [hep-ph]].
  %%CITATION = PHRVA,D78,034025;%%

\bibitem{Boffi:2009sh}
  S.~Boffi, A.~V.~Efremov, B.~Pasquini and P.~Schweitzer,
  %``Azimuthal spin asymmetries in light-cone constituent quark models,''
  Phys.\ Rev.\  D {\bf 79}, 094012 (2009).
  %[arXiv:0903.1271 [hep-ph]].
  %%CITATION = PHRVA,D79,094012;%%

\bibitem{Pasquini:2010pa}
  B.~Pasquini and C.~Lorcé,
  %``Modeling the transverse momentum dependent parton distributions,''
  arXiv:1008.0945 [hep-ph].
  %%CITATION = ARXIV:1008.0945;%%

\bibitem{Lorce:2011dv}
  C.~Lorc\'e, B.~Pasquini and M.~Vanderhaeghen,
  %``Unified framework for generalized and transverse-momentum dependent parton
  %distributions within a 3Q light-cone picture of the nucleon,''
  arXiv:1102.4704 [hep-ph].
  %%CITATION = ARXIV:1102.4704;%%





\bibitem{Pasquini:2010af}
  B.~Pasquini and F.~Yuan,
  %``Sivers and Boer-Mulders functions in Light-Cone Quark Models,''
  Phys.\ Rev.\  D {\bf 81}, 114013 (2010).
  %%[arXiv:1001.5398 [hep-ph]].
  %%CITATION = PHRVA,D81,114013;%%

\bibitem{Brodsky:2010vs}
  S.~J.~Brodsky, B.~Pasquini, B.~W.~Xiao and F.~Yuan,
  %``Phases of Augmented Hadronic Light-Front Wave Functions,''
  Phys.\ Lett.\  B {\bf 687}, 327 (2010).
  %%[arXiv:1001.1163 [hep-ph]].
  %%CITATION = PHLTA,B687,327;%%

\bibitem{Bacchetta:2008af}
  A.~Bacchetta, F.~Conti and M.~Radici,
  %``Transverse-momentum distributions in a diquark spectator model,''
  Phys.\ Rev.\  D {\bf 78}, 074010 (2008).
  %[arXiv:0807.0323 [hep-ph]].
  %%CITATION = PHRVA,D78,074010;%%

\bibitem{Avakian:2010br}
  H.~Avakian, A.~V.~Efremov, P.~Schweitzer and F.~Yuan,
  %``The transverse momentum dependent distribution functions in the bag
  %model,''
  Phys.\ Rev.\  D {\bf 81},  074035 (2010).
  %[arXiv:1001.5467 [hep-ph]].
  %%CITATION = PHRVA,D81,074035;%%

\bibitem{Avakian:2008dz}
  H.~Avakian, A.~V.~Efremov, P.~Schweitzer and F.~Yuan,
  %``Transverse momentum dependent distribution function $h_{1T}^\perp$ and the
  %single spin asymmetry $A_{UT}^{\sin(3\phi-\phi_S)}$,''
  Phys.\ Rev.\  D {\bf 78}, 114024 (2008).
  %[arXiv:0805.3355 [hep-ph]].
  %%CITATION = PHRVA,D78,114024;%%



\bibitem{Efremov:2010mt}
  A.~V.~Efremov, P.~Schweitzer, O.~V.~Teryaev and P.~Zavada,
  %``The relation between TMDs and PDFs in the covariant parton model
  %approach,''
  arXiv:1012.5296 [hep-ph].
  %%CITATION = ARXIV:1012.5296;%%



\bibitem{Meissner:2007rx}
  S.~Meissner, A.~Metz and K.~Goeke,
  %``Relations between generalized and transverse momentum dependent parton
  %distributions,''
  Phys.\ Rev.\  D {\bf 76}, 034002 (2007).
  %[arXiv:hep-ph/0703176].
  %%CITATION = PHRVA,D76,034002;%%
\bibitem{Wakamatsu:2009fn}
  M.~Wakamatsu,
  %``Transverse momentum distributions of quarks in the nucleon from the Chiral
  %Quark Soliton Model,''
  Phys.\ Rev.\  D {\bf 79}, 094028 (2009).
  %[arXiv:0903.1886 [hep-ph]].
  %%CITATION = PHRVA,D79,094028;%%


\bibitem{Schweitzer:2001sr}
  P.~Schweitzer, D.~Urbano, M.~V.~Polyakov, C.~Weiss, P.~V.~Pobylitsa
  and K.~Goeke,
  %%{\it Transversity distributions in the nucleon in the large-N$_c$ limit},
  Phys.\ Rev. D {\bf 64}, 034013 (2001).
  %%[\href{http://arxiv.org/abs/hep-ph/0101300}{{\tt hep-ph/0101300 }}].
  %%[arXiv:hep-ph/0101300].
  %%CITATION = HEP-PH 0101300;%%




\bibitem{Mulders:1995dh}
  P.~J.~Mulders and R.~D.~Tangerman,
  %``The complete tree-level result up to order 1/Q for polarized
  %deep-inelastic leptoproduction,''
  Nucl.\ Phys.\  B {\bf 461}, 197 (1996) 
  [Erratum-ibid.\  B {\bf 484}, 538 (1997)].
  %[arXiv:hep-ph/9510301].
  %%CITATION = NUPHA,B461,197;%%
  
\bibitem{Boer:1997nt}
  D.~Boer and P.~J.~Mulders,
  %``Time-reversal odd distribution functions in leptoproduction,''
  Phys.\ Rev.\  D {\bf 57}, 5780 (1998).
  %[arXiv:hep-ph/9711485].
  %%CITATION = PHRVA,D57,5780;%%
  
\bibitem{Goeke:2005hb}
  K.~Goeke, A.~Metz and M.~Schlegel,
  %``Parameterization of the quark-quark correlator of a spin-1/2 hadron,''
  Phys.\ Lett.\  B {\bf 618}, 90 (2005).
  %[arXiv:hep-ph/0504130].
  %%CITATION = PHLTA,B618,90;%%
  

\bibitem{Bomhof:2004aw}
  C.~J.~Bomhof, P.~J.~Mulders and F.~Pijlman,
  %``Gauge link structure in quark quark correlators in hard processes,''
  Phys.\ Lett.\  B {\bf 596}, 277 (2004).
  %[arXiv:hep-ph/0406099].
  %%CITATION = PHLTA,B596,277;%%

\bibitem{Sivers:1989cc}
  D.~W.~Sivers,
  %``Single Spin Production Asymmetries From The Hard Scattering Of Point-Like
  %Constituents,''
  Phys.\ Rev.\  D {\bf 41}, 83 (1990).
  %%CITATION = PHRVA,D41,83;%%

\bibitem{Diehl:2000xz}
  M.~Diehl, T.~Feldmann, R.~Jakob and P.~Kroll,
  %``The overlap representation of skewed quark and gluon distributions,''
  Nucl.\ Phys.\  B {\bf 596}, 33 (2001)
  [Erratum-ibid.\  B {\bf 605}, 647 (2001)].
  %[arXiv:hep-ph/0009255].
  %%CITATION = NUPHA,B596,33;%%

\bibitem{Brodsky:2000xy}
  S.~J.~Brodsky, M.~Diehl and D.~S.~Hwang,
  %``Light-cone wavefunction representation of deeply virtual Compton
  %scattering,''
  Nucl.\ Phys.\  B {\bf 596}, 99 (2001).
  %[arXiv:hep-ph/0009254].
  %%CITATION = NUPHA,B596,99;%%


\bibitem{Hagler:2009mb}
  Ph.~Hagler, B.~U.~Musch, J.~W.~Negele and A.~Schafer,
  %``Intrinsic quark transverse momentum in the nucleon from lattice QCD,''
  Europhys.\ Lett.\  {\bf 88}, 61001 (2009).
  %[arXiv:0908.1283 [hep-lat]].
  %%CITATION = EULEE,88,61001;%%

\bibitem{Musch:2010ka}
  B.~U.~Musch, P.~Hagler, J.~W.~Negele and A.~Schafer,
  %``Exploring quark transverse momentum distributions with lattice QCD,''
  arXiv:1011.1213 [hep-lat];
  %%CITATION = ARXIV:1011.1213;%%



\bibitem{Efremov:2002qh}
  A.~V.~Efremov and P.~Schweitzer,
  %``The chirally-odd twist-3 distribution (e**a)(x),''
  JHEP {\bf 0308}, 006 (2003).
  %[arXiv:hep-ph/0212044].
  %%CITATION = JHEPA,0308,006;%%


\bibitem{Lorce:2006nq}
  C.~Lorcé,
  %``Improvement of the Theta+ width estimation method on the light cone,''
  Phys.\ Rev.\  D {\bf 74}, 054019 (2006).
  %[arXiv:hep-ph/0603231].
  %%CITATION = PHRVA,D74,054019;%%



\bibitem{Lorce:2007as}
  C.~Lorcé,
  %``Baryon vector and axial content up to the 7Q component,''
  Phys.\ Rev.\  D {\bf 78}, 034001 (2008).
  %[arXiv:0708.3139 [hep-ph]].
  %%CITATION = PHRVA,D78,034001;%%


\bibitem{Melosh:1974cu}
  H.~J.~Melosh,
  %``Quarks: Currents and constituents,''
  Phys.\ Rev.\  D {\bf 9}, 1095 (1974).
  %%CITATION = PHRVA,D9,1095;%%

\bibitem{Ahluwalia:1993xa}
  D.~V.~Ahluwalia and M.~Sawicki,
  %``Front form spinors in the Weinberg-Soper formalism and generalized Melosh
  %transformations for any spin,''
  Phys.\ Rev.\  D {\bf 47}, 5161 (1993).
  %[arXiv:nucl-th/9603019].
  %%CITATION = PHRVA,D47,5161;%%


  
\bibitem{Petrov:2002jr}
  V.~Y.~Petrov and M.~V.~Polyakov,
  %``Light cone nucleon wave function in the quark soliton model,''
  arXiv:hep-ph/0307077.
  %%CITATION = HEP-PH/0307077;%%

\bibitem{Diakonov:2005ib}
  D.~Diakonov and V.~Petrov,
  %``Estimate of the Theta+ width in the relativistic mean field
  %approximation,''
  Phys.\ Rev.\  D {\bf 72}, 074009 (2005).
  %[arXiv:hep-ph/0505201].
  %%CITATION = PHRVA



\bibitem{Brodsky:2000ii}
  S.~J.~Brodsky, D.~S.~Hwang, B.~Q.~Ma and I.~Schmidt,
  %``Light cone representation of the spin and orbital angular momentum of
  %relativistic composite systems,''
  Nucl.\ Phys.\  B {\bf 593}, 311 (2001).
  %[arXiv:hep-th/0003082].
  %%CITATION = NUPHA,B593,311;%%
\bibitem{Ellis:2008in}
  J.~R.~Ellis, D.~S.~Hwang and A.~Kotzinian,
  %``Sivers Asymmetries for Inclusive Pion and Kaon Production in Deep-Inelastic
  %Scattering,''
  Phys.\ Rev.\  D {\bf 80}, 074033.(2009).
  %[arXiv:0808.1567 [hep-ph]].
  %%CITATION = PHRVA,D80,074033;%%

\bibitem{Gross:2006fg}
  F.~Gross, G.~Ramalho and M.~T.~Pena,
  %``A Pure S-wave covariant model for the nucleon,''
  Phys.\ Rev.\  C {\bf 77}, 015202 (2008).
  %%[arXiv:nucl-th/0606029].
  %%CITATION = PHRVA,C77,015202;%%

\bibitem{Ramalho:2008ra}
  G.~Ramalho, M.~T.~Pena and F.~Gross,
  %``A Covariant model for the nucleon and the Delta,''
  Eur.\ Phys.\ J.\  A {\bf 36}, 329 (2008).
  %% [arXiv:0803.3034 [hep-ph]].
  %%CITATION = EPHJA,A36,329;%%

\bibitem{Gross:2008zza}
  F.~Gross, G.~Ramalho and M.~T.~Pena,
  %``Fixed-axis polarization states: covariance and comparisons,''
  Phys.\ Rev.\  C {\bf 77}, 035203 (2008).
  %%CITATION = PHRVA,C77,035203;%%

%\bibitem{Jaffe:1974nj}
%  R.~L.~Jaffe,
%  %``Deep Inelastic Structure Functions In An Approximation To The Bag Theory,''
%  Phys.\ Rev.\  D {\bf 11} (1975) 1953.
%  %%CITATION = PHRVA,D11,1953;%%

%\bibitem{Jaffe:1975yc}
%  R.~L.~Jaffe and A.~Patrascioiu,
%  %``Light Cone Structure In The Cavity Approximation To The Bag Theory,''
%  Phys.\ Rev.\  D {\bf 12} (1975) 1314.
%  %%CITATION = PHRVA,D12,1314;%%

%\bibitem{Weigel:1996kw}
%  H.~Weigel, L.~P.~Gamberg and H.~Reinhardt,
%  %``Nucleon Structure Functions from a Chiral Soliton,''
%  Phys.\ Lett.\  B {\bf 399} (1997) 287
%  [arXiv:hep-ph/9604295].
%  %%CITATION = PHLTA,B399,287;%%

%\bibitem{Gamberg:1997qk}
%  L.~P.~Gamberg, H.~Reinhardt and H.~Weigel,
%  %``Nucleon structure functions from a chiral soliton in the infinite  momentum
%  %frame,''
%  Int.\ J.\ Mod.\ Phys.\  A {\bf 13} (1998) 5519
%  [arXiv:hep-ph/9707352].
%  %%CITATION = IMPAE,A13,5519;%%

%\bibitem{Christov:1995hr}
%  C.~V.~Christov, A.~Z.~Gorski, K.~Goeke and P.~V.~Pobylitsa,
%  %``Electromagnetic Form-Factors Of The Nucleon In The Chiral Quark Soliton
%  %Model,''
%  Nucl.\ Phys.\  A {\bf 592}, 513 (1995).
%  %[arXiv:hep-ph/9507256].
%  %%CITATION = NUPHA,A592,513;%%



\end{thebibliography}
\end{document}